\documentclass[10pt,conference]{IEEEtran}

\usepackage[]{hyperref}
\usepackage{multirow}
\usepackage{amsmath}
\usepackage{makecell}
\usepackage{booktabs}
\usepackage{algorithmic}
\usepackage{graphicx}
\usepackage{textcomp}
\usepackage{xcolor}
\usepackage{tikz}
\usepackage{subcaption}
\usepackage{braket}
\AtBeginDocument{%
  }

\usepackage{qcircuit}

\begin{document}
\title{
Leveraging Hardware Power through Optimal Pulse Profiling for Each Qubit Pair
}
\author{\IEEEauthorblockN{
Yuchen Zhu\IEEEauthorrefmark{2},
Jinglei Cheng\IEEEauthorrefmark{3},
Boxi Li\IEEEauthorrefmark{4},
Yidong Zhou\IEEEauthorrefmark{2},
Yufei Ding\IEEEauthorrefmark{5},
Zhiding Liang\IEEEauthorrefmark{2}\\
\IEEEauthorblockA{
\IEEEauthorrefmark{2}
Rensselaer Polytechnic Institute, Troy, NY, USA\\
\IEEEauthorrefmark{3}
University of Pittsburgh, Pittsburgh, PA, USA\\
\IEEEauthorrefmark{4}
Forschungszentrum Jülich, Jülich, Germany\\
\IEEEauthorrefmark{5}
University of California San Diego, San Diego, CA, USA}}
(liangz9@rpi.edu)
}


\maketitle
\begin{abstract}
In the scaling development of quantum computers, the calibration process emerges as a critical challenge. 
Existing calibration methods, utilizing the same pulse waveform for two-qubit gates across the device, overlook hardware differences among physical qubits and lack efficient parallel calibration.
In this paper, we enlarge the pulse candidates for two-qubit gates to three pulse waveforms, and introduce a fine-grained calibration protocol. 
In the calibration protocol, three policies are proposed to profile each qubit pair with its optimal pulse waveform.
Afterwards, calibration subgraphs are introduced to enable parallel calibraton through identifying compatible calibration operations.
The protocol is validated on real machine with up to 127 qubits.
Real-machine experiments demonstrates a minimum gate error of 0.001 with a median error of 0.006 which is 1.84 $\times$ reduction compared to default pulse waveform provided by IBM.
On device level, a double fold increase in quantum volume as well as  2.3 $\times$ reduction in error per layered gate are achieved.
The proposed protocol leverages the potential current hardware and could server as an important step toward fault-tolerant quantum computing.


\end{abstract}
\section{Introduction}
\label{introduction}

Recent advances in quantum computing have demonstrated impressive scaling in qubit numbers, with leading platforms now reaching hundreds of physical qubits~\cite{abughanem2024ibm, DWaveAdvantage}. 
However, achieving high-fidelity quantum operations on these large-scale systems remains a critical challenge~\cite{xu2020high, bao2022fluxonium,tannu2017taming,das2021adapt}. 
While single-qubit gate fidelities often exceed 99.9\%~\cite{huang_fidelity_2019,bruzewicz2019trapped}, two-qubit gate fidelities on multi-qubit, fixed-frequency superconducting chips typically lie around 99\%~\cite{gonzalez2022silicon,vinet2018towards}, significantly higher than the 0.1\% error rates on specialized two-qubit devices. 
As one of the leading quantum computing platforms, superconducting quantum systems also face challenges in maintaining high-fidelity operations as the system scales up, which signifies the importance of a good calibration protocol.

\begin{figure}
    \centering
    \includegraphics[width=\linewidth]{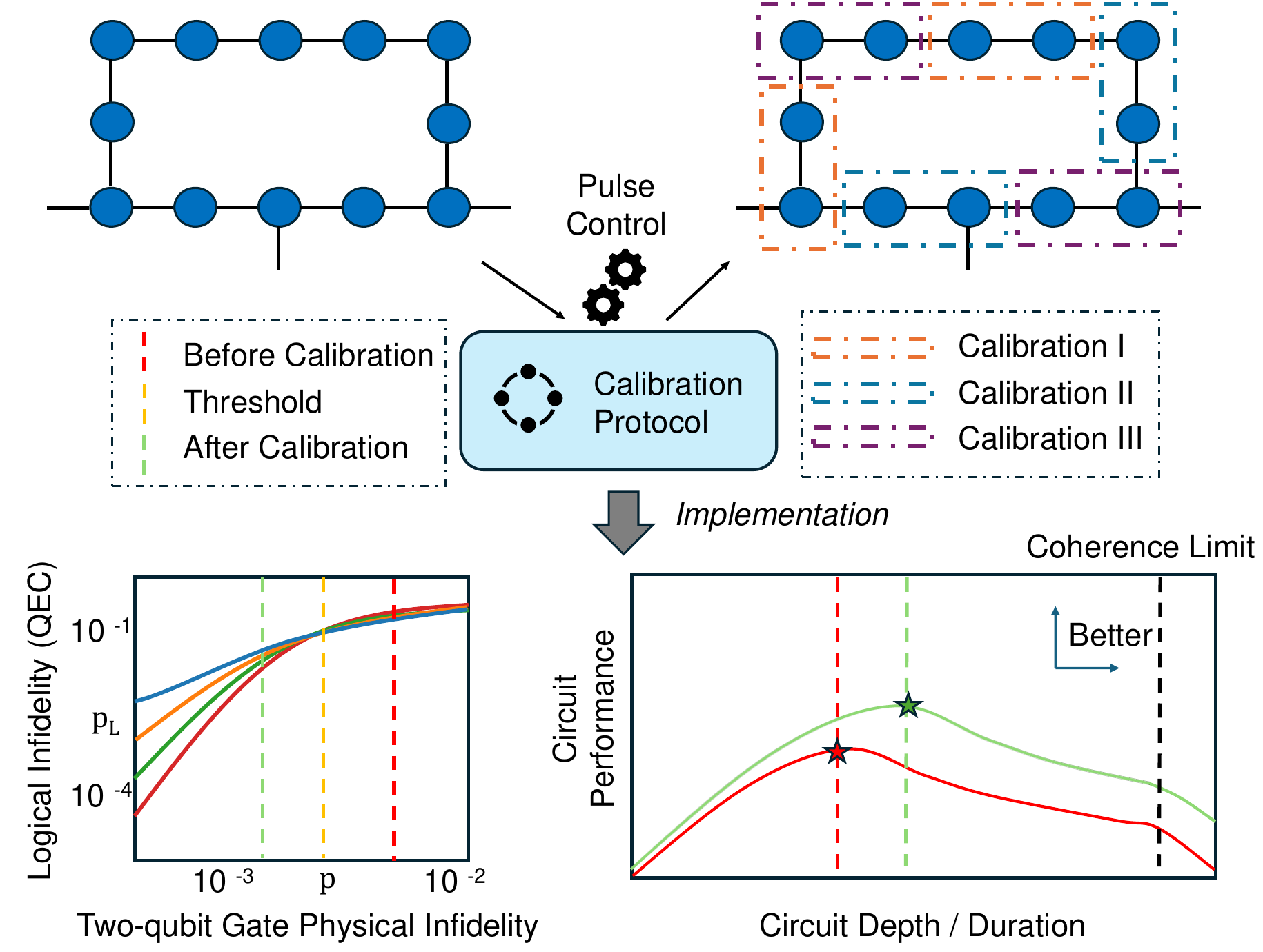}
    \caption{Overview of the proposed hardware-aware calibration protocol and its impact.}
    \label{fig:teaser}
    \vspace{-3mm}
\end{figure}

Superconducting quantum computers operate at extremely low temperatures in high-vacuum environments to maintain quantum coherence~\cite{Krantz2019Quantum}. 
These systems rely on microwave pulse control for qubit manipulation and measurements. 
The microwave signals define the system Hamiltonian and the qubit evolution, while the reflected signals of measurement pulses enable quantum state measurements. 
The quality of quantum operations depends on the precise calibration of these control systems~\cite{chen2023quantum,kanazawa2023qiskit}.
Precise control involves two key aspects: generating accurate microwave pulses and performing regular calibrations to compensate for system drift and external perturbations. 
The control parameters include pulse amplitudes, frequencies, and durations, which must be carefully calibrated based on measurements of current qubit properties~\cite{Koch2007Chargeinsensitive,gokhale2020optimized,vandersypen2004nmr}. 
To maintain optimal operation conditions and keep error rates within acceptable ranges, quantum hardware providers like IBM perform calibration routines at both hourly and daily intervals~\cite{IBMQuantumCalibrationJobs}. 
Regular calibration ensures that qubits perform the designed operations. 
While individual gate errors may appear insignificant in isolation, their cumulative effect becomes significant as circuit depth increases and the number of gates grows. 
As these errors accumulate, they can significantly degrade the overall fidelity of the quantum circuit, and eventually make the execution of quantum algorithms unreliable~\cite{koczor2024probabilistic, wang2022torchquantum,ravi2022vaqem,xu2023exploration,huang2021logical}. 

Calibration is also critical for Quantum Error Correction (QEC) since when effective calibration is applied to reduce the physical error rates below a critical threshold, QEC can exponentially suppress logical errors by increasing the code distance~\cite{chiaverini2004realization, sivak2023real}. 
Recent advancements, such as below-threshold operation of surface codes~\cite{noh2022low} and hardware-efficient schemes like bosonic cat qubits~\cite{xu2023autonomous}, have shed light on the potential of QEC in the near future. 
These advances have driven extensive experimental efforts to optimize QEC protocols in real quantum systems.
Related works~\cite{benito2024comparative, kubica2022single, schindler2011experimental, moussa2011demonstration} demonstrate that achieving high-fidelity operations for practical QEC demands precise calibration. 
However, traditional calibration methods face significant challenges. 
First, they often ignore differences in qubits' properties. Second, the extended duration of calibration—spanning hours or days~\cite{IBMQuantumCalibrationJobs}—introduces system drift that affects the system performance. 
Finally, existing approaches emphasize a trade-off between calibration time and fidelity but neglect hardware-specific optimizations and the strict timing requirements of QEC protocols. 
These limitations are particularly severe in calibrating cross-resonance (CR) gates which are essential for two-qubit operations in modern quantum processors.
As QEC becomes a cornerstone of fault-tolerant quantum computing, it imposes rigorous demands on gate fidelities and control accuracies.

A key challenge in CR gates arises from unwanted excitations of the control qubit due to off-resonant drives. 
The standard derivative removal by adiabatic gate (DRAG) technique addresses this issue for single transitions~\cite{motzoi2009simple,gambetta2011analytic}, but quantum systems often exhibit multiple transition pathways that need simultaneous control. 
Multi-derivative DRAG offers a solution by applying recursive corrections to suppress multiple transitions simultaneously~\cite{Li2024Experimental}. 
This method adds derivative terms to the control pulse that target specific energy transitions between quantum states. 
While this approach can enhance gate fidelity, its effectiveness depends on system parameters such as the frequency difference between qubits. 
These implementation considerations motivate the need for a systematic approach to optimize calibration across multiple qubit pairs simultaneously.

To optimize the calibration process across qubit pairs, we introduce a novel calibration protocol with three different calibration policies:\textbf{(1) Brute-force Clustering}, where qubit pairs are grouped based on key physical properties, representative pairs are selected for calibration, and the optimized Echoed Cross-Resonance (ECR) gates' waveforms are generalized; \textbf{(2) Topology-oriented Representative}, which leverages the regular patterns in heavy-hex lattice topologies by classifying qubit pairs according to their positions within unit cells, calibrates representatives for each position, and generalizes the optimized waveforms; and \textbf{(3) Hardware-oriented Policy}, which incorporates system knowledge and hardware limitations, such as frequency detuning relationships and decoherence times, to select the most efficient and practical waveform strategies for each qubit pair. 

Our policies also offers a practical balance between calibration accuracy and computational efficiency, which is particularly important for scaling up quantum processors. 
However, successful local optimization of qubit pairs is only the first step toward processor-wide calibration.
We address processor-wide calibration by treating the quantum processor as an undirected graph G, where nodes represent qubits and edges represent coupled qubit pairs. 
To minimize calibration time, we develop a parallel calibration protocol that partitions the coupling graph into calibration subgraphs. Each subgraph consists of edges that can be calibrated simultaneously without mutual interference, achieved by maintaining a minimum distance of two between concurrent calibrations. 
This parallelization strategy significantly accelerates the calibration process while preserving calibration accuracy. For instance, in a 127-qubit heavy-hex architecture, our approach enables simultaneous calibration of up to 38 qubit pairs distributed across five subgraphs, substantially reducing the total calibration time compared to sequential methods. 

We illustrate the brief overview of our protocol in Figure~\ref{fig:teaser}.
Our protocol optimizes calibration policies and assigns different pulse envelopes to qubit pairs as the colored blocks in the top-right of Figure~\ref{fig:teaser}. 
The bottom-left of Figure~\ref{fig:teaser} shows the effectiveness of quantum error correction at various code distances, represented by different colored curves. 
Our protocol achieves two key improvements: it reduces two-qubit gate physical error rates below the error correction threshold, enabling effective quantum error correction and significantly improving logical error rates; 
and it optimizes system performance by reducing calibration overhead, shortening circuit duration, and enhancing overall circuit fidelity compared to conventional calibration approaches, as shown in the bottom-right of Figure~\ref{fig:teaser}.

In this work, we introduce a novel fine-grained calibration protocol with three major contributions:
\begin{itemize}
   \item A novel calibration protocol with multiple components: 1) Three calibration policies to assign waveform candidate to qubit pairs according to their profiling results. 2) A graph-based parallelization algorithm for scaling up the protocol that performs the calibration simultaneously.
   
   \item The first large-scale implementation of multi-derivative DRAG and direct CR operations on real quantum machines, which advances quantum control techniques in practice.
   
   \item Extensive experimental results with up to 127 qubits that show a $1.84 \times$ reduction in terms of the medium of the two-qubit gate error rate, $1.26 \times$ reduction in pulse duration, up to $25 \times$ reduction in calibration overhead, double of the quantum volume, and up to $2.3 \times$ reduction in error per layered gate.
\end{itemize}

\section{Background}
\label{background}

\subsection{Calibration and Characterization in Quantum Gates}

In quantum computing, qubits and quantum gates exhibit extreme sensitivity to external disturbances~\cite{preskill2018quantum}. 
Calibration and characterization thus serve as essential procedures in quantum hardware experiments. 
For superconducting qubits, quantum gates operate through precise control pulses, which require careful tuning of multiple parameters: drive frequency, amplitude, phase, and duration~\cite{liang2024napa}. 
Each parameter affects the qubit's dynamics directly. For example, small amplitude deviations can cause over-rotations or under-rotations of qubits. These errors accumulate and degrade overall circuit performance and fidelity as operations scale.
Therefore, amplitude adjustments must be performed through iterative, fine-grained calibration routines. This process requires amplitude tuning in small increments to achieve the desired qubit state transition, which aligns actual gate behavior more closely with the intended ideal operation. 
Moreover, these parameters need continuous optimization to maintain stability over time, particularly in the presence of inherent hardware variations such as frequency drifts and crosstalk. 
To optimize the calibration of qubits, we need to understand the properties of quantum gates on the qubits.
Characterization techniques reveal crucial insights into the performance of quantum gates in physical devices~\cite{patel2020experimental}. Quantum Process Tomography (QPT) is an approach that constructs the entire quantum process matrix for a certain quantum gate~\cite{mohseni2008quantum}. This method requires qubit preparation in various input states, gate application, and measurement of output states to characterize the gate's behavior fully. 
However, QPT's bad scalability and it's not applicable to larger systems.
Randomized Benchmarking (RB)~\cite{knill2008randomized} presents a more efficient alternative for characterization. This statistical approach determines average gate error rates through random gate sequences applied to qubits. Modern implementations often incorporate interleaved randomized benchmarking~\cite{magesan2012efficient}, where a specific gate of interest alternates with random Clifford operations, enabling characterization of individual gate performance within the broader context of circuit operation.

\subsection{Echoed CR}
The Echoed Cross Resonance (Echoed CR) pulse~\cite{alexander2020qiskit} is an existing implementation method for two-qubit gates in superconducting quantum computing, particularly for the CNOT gate implementation.
The standard CR pulse has multiple Hamiltonian terms: $ ZX, ZZ, ZI, IX, IY $, and others. 
The Echoed CR pulse is built upon the CR pulse by adding an ``echo'' mechanism to reduce phase errors and unwanted couplings during gate operation.
This echoing procedure eliminates unwanted Hamiltonian terms and preserves the desired $ZX$ interaction. 
The result implements an effective gate operation of $\frac{1}{\sqrt{2}} \left( IX - XY \right)$. 
Therefore, to calibrate the Echoed CR pulse, we need to calibrate the CR pulse within it.
Unwanted $ZZ$ interaction that causes phase errors can be addressed by $ZY$ DRAG calibration~\cite{mckay2018qiskit}. 
This technique modifies the control pulse through a standard Gaussian pulse with an additional Gaussian derivative component and lifting.
The calibration process focuses on the selective cancellation of remaining $ZY, IX, IY$ terms through iterative adjustments of CR and target qubit drive pulses. 
The system performs two tomography experiments per calibration round to refine the pulse parameters. 
Hamiltonian tomography~\cite{sheldon2016procedure} is adopted to measure the target qubit's response with various time intervals. 
The data from Hamiltonian tomography enables the decomposition of the Hamiltonian into Pauli terms:
\begin{equation}
\hat{H}(\Omega_{\text{CR}}, \Omega_{\text{T}}) = \nu_{ZX} \hat{Z} \hat{X} + \nu_{ZY} \hat{Z} \hat{Y} + \nu_{IX} \hat{I} \hat{X} + \nu_{IY} \hat{I} \hat{Y} + \nu_{ZI} \hat{Z} \hat{I}..
\label{eq:two-qubit CR ham}
\end{equation}
The suppression of $ZZ$ effects requires precise $IY$ DRAG calibration. 
This process involves sampling three different $IY$ DRAG amplitudes to identify zero points of the $ZZ$ coupling strength through linear fitting. 
The measured $IZ$ coefficient represents the detuning strength between drives. 
Proper calibration of the $IY$ DRAG reduces $ZZ$ coupling, which would otherwise affect CR pulse fidelity.
The final implementation uses an echoed pulse sequence. 
This sequence consists of a primary CR pulse followed by a reversed CR pulse. 
For an echoed CNOT gate, each CR pulse must last one-eighth of the period, as the target qubit rotates in opposite directions based on the control qubit state. This precise timing ensures accurate 90-degree ZX rotation.

\subsection{Direct CR}
\label{subsec-cali}

Direct CR~\cite{chow2011simple} is more expensive in terms of calibration cost against Echoed CR because of it based on Echoed CR with additional technologies involved.  

\begin{figure}[h!]
    \centering
    \begin{subfigure}[t]{\linewidth}
        \centering
        \includegraphics[width=0.85\textwidth]{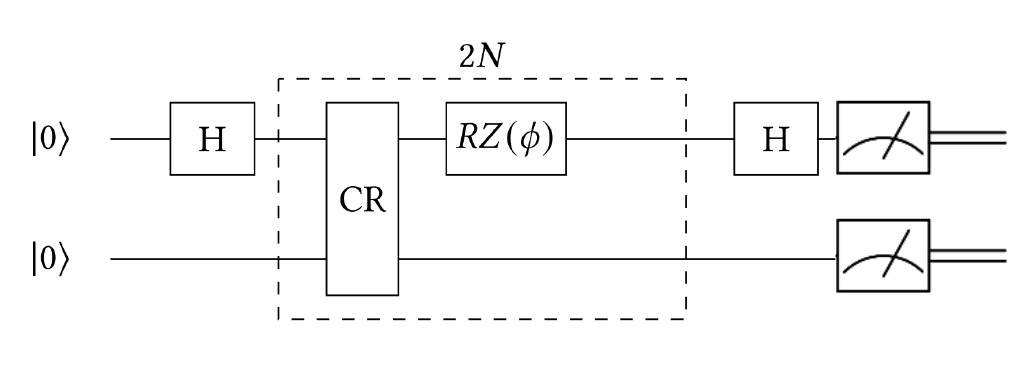} 
        \caption{}
        \label{fig:directcr1}
    \end{subfigure}
    
    \vspace{1em} %

    \begin{subfigure}[t]{\linewidth}
        \centering
        \includegraphics[width=\textwidth]{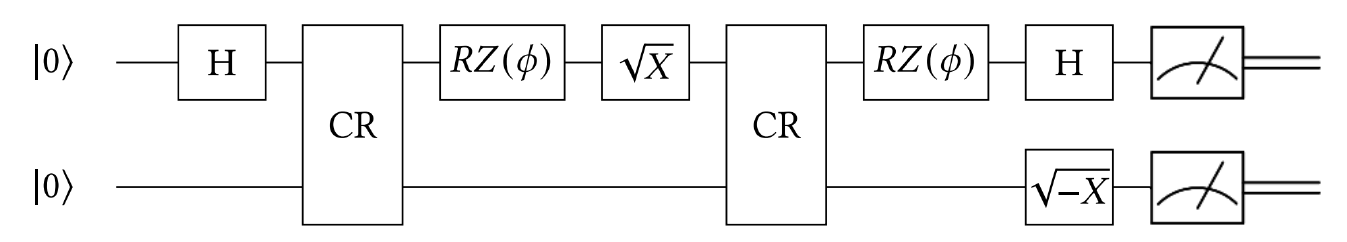} 
        \caption{}
        \label{fig:directcr2}
    \end{subfigure}
    
    \caption{(a) Calibration circuit for Z phase for the direct CR. (b) Verification circuit for the direct CR.}
    \label{fig:directcr}
\end{figure}

The calibration process begins with adjustment on target qubit signals to satisfy the condition $\nu_{IX} = \nu_{ZX}$. 
This adjustment ensures the target qubit rotates only when the control qubit exists in the $\ket{1}$ state. 
The adjustment is implemented through the iterative calibration process described in the echoed CR calibration.
The next phase addresses the problem of phase shift on control qubit, which is introduced by the Stark effect with the CR pulse. 
This calibration step is the key difference from echoed CR calibration. 
In summary, the Echoed CR pulse automatically cancels phase shifts, but direct CR requires specific phase calibration.
The phase calibration process employs two specific circuit configurations, as shown in Figure \ref{fig:directcr1} and \ref{fig:directcr2}. 
The circuit in Figure \ref{fig:directcr1} implements a sequence of $2N$ uncalibrated CR pulses, each paired with an $R_Z(\phi)$ rotation on the control qubit. 
Hadamard gates bracket this sequence to measure the accumulated phase shifts. 
The circuit returns to its initial state only when the CR pulses combined with the rotation produce a standard CNOT with a 0-degree or 180-degree phase shift. 
Figure \ref{fig:directcr2} presents a verification circuit that confirms the implementation of a proper CNOT gate.

\begin{figure*}[t]
    \centering
    \includegraphics[width=\linewidth]{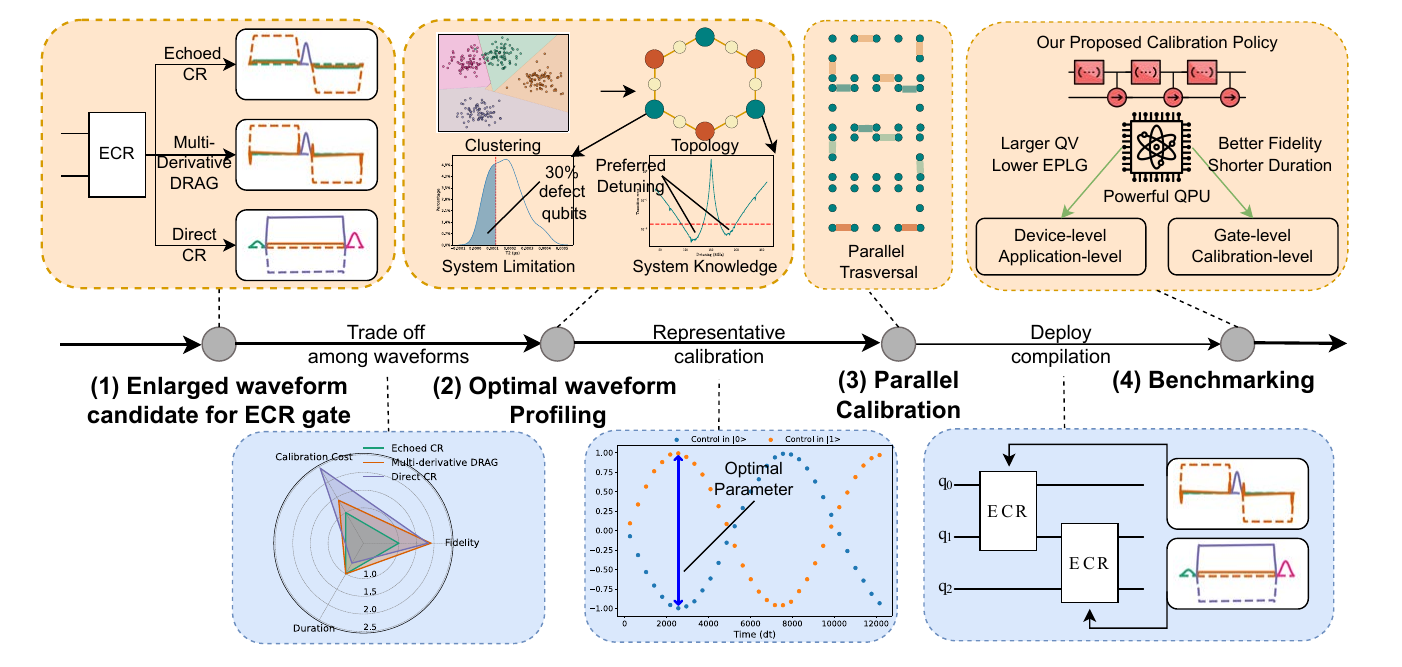}
    \caption{Overview of the proposed calibration protocol design. With enlarged pulse candidates for ECR gate, optimal waveform profiling is achieved by the proposed calibration policies. Parallel calibration is adopted to accelerate the calibration process. The effectiveness of the protocol is examined from gate-level, calibration-level, device-level, and application-level.}
    \label{fig:overview}
\end{figure*}

\section{Motivation}
\label{motivation}

Existing calibration methods face two major challenges: hardware inconsistencies and high calibration costs. While recent advancements propose solutions through noise-aware optimization, adaptive protocols, and real-time feedback, these approaches often introduce significant overhead in both computation and time resources. This overhead can substantially reduce quantum processor availability for practical workloads. Therefore, an improved calibration protocol must achieve three key objectives: (1) address hardware-specific variations, (2) maintain high fidelity with reasonable latency, and (3) minimize calibration costs while ensuring system performance. 

\subsection{Hardware Variability in Superconducting Quantum Systems}

\label{superconductingquantumhardwarelimitations}

Hardware limitations fundamentally restrict the performance of superconducting quantum computers. These constraints result in weak qubit interaction strength and low quantum gate fidelity, which make precise calibration technology essential for reliable operation.
The significant differences in relaxation time (T\textsubscript{1}) and dephasing time (T\textsubscript{2}) between qubit pairs create a critical challenge. 
Short decoherence times lead to rapid quantum information loss in qubits, which directly affects the fidelity of two-qubit operations. 
This inconsistency requires individual calibration protocols for each qubit pair and adds substantial complexity to system maintenance.
Furthermore, fabrication variations and material differences cause wide frequency detuning values across qubit pairs. 
These inherent hardware variations mean that standardized calibration methods and uniform control pulses \textbf{cannot} achieve optimal performance across all qubit pairs, which results in reduced overall gate fidelity. 
This hardware-specific behavior demands more sophisticated, tailored calibration approaches.
We address this problem in Section~\ref{calibrationpolicydesign}.

\subsection{Missing Performance Metrics in Traditional Calibration}

Traditional calibration methods rely on single-gate fidelity as the primary performance metric, which provides an incomplete picture of quantum ~\textbf{system} performance. 
Advanced compilation techniques~\cite{tannu2022hammer,tomesh2021quantum,tan2020optimal,liu2022not} have revealed that optimizing individual gate fidelity alone fails to capture critical system-level effects, such as qubit idle times and decoherence during circuit execution. 
Quantum Volume addresses these limitations by providing a metric that includes both qubit quality and system performance. 
However, current calibration techniques continue to focus on optimizing isolated metrics like gate fidelity or latency, without incorporating metrics like Quantum Volume. This narrow focus on performance assessment leads to suboptimal calibration strategies that fail to capture the true capabilities of quantum devices.
We demonstrate gate-level, calibration-level, device-level and application-level improvements of our protocol in the Section~\ref{evaluation}.

\subsection{High-overhead and Infrequent Calibration}
\label{concernsincalibrationcosts}

Current generalized calibration approaches face significant limitations due to the high costs associated with the calibration process. 
These costs lead to infrequent calibration, which allows system drift and errors to accumulate over time. 
The problem is especially severe for two-qubit gates, which are essential for advanced quantum algorithms and error correction but require longer calibration times to maintain coherence.
There is an inherent trade-off between more frequent calibration to improve gate fidelity and the system downtime this calibration requires. 
On IBM quantum hardware, two-qubit gates need long calibration cycles to reach acceptable fidelity levels, and this results in extended downtimes. 
During these periods, system properties can drift, and this causes unexpected errors even in simpler single-qubit gates. Current calibration approaches do not perform calibration often enough to sufficiently mitigate these issues.

Experiments on IBM hardware demonstrate the consequences of insufficient calibration. Measurements showed error rates for single-qubit gates that reached as high as $10^{-2}$ when trying to excite qubits to the $\ket{1}$ state. These high error rates highlight the impact of infrequent calibration on gate fidelity.
As circuits grow in size, errors from two-qubit gates and drifting system properties accumulate, and this significantly affects the computational results.
IBM's current calibration standards focus on weekly full calibration of only a limited number of qubit pairs. 
Daily measurements include phase calibrations for just a few pairs. 
This approach allows substantial system drift to occur, as the minor fidelity improvements from more frequent calibration are considered unjustified given the high calibration costs.
This significantly limits gate fidelity and computational accuracy, especially as quantum circuits become larger. 
We address the challenge of large-overhead calibration by parallel calibration as described in Section~\ref{graphtraverse}.

\section{Protocol}
\label{method}
In this section, we are introducing our proposed calibration protocol. The performance of current quantum hardware is leveraged and characterized through 4 steps, as depicted in Figure~\ref{fig:overview}.

\noindent \textbf{Enlargement of candidate waveform.} Three distinct waveforms are developed, each capable of implementing the same basis two-qubit gate on current quantum hardware. These waveforms offer trade-offs among gate fidelity, calibration cost, and gate duration, providing flexibility in optimizing quantum operations based on system constraints.\\
\textbf{Optimal pulse profiling.} In this step, multiple policies are employed, aiming at profiling each qubit pair with its own optimal pulse waveform. The policies are designed upon the physical properties, topology similarity and system knowledge.\\
\textbf{Parallel Calibration.} After the optimal pulse is decided, a parallel graph traversal is implemented to calibrate all qubit pairs with limited time.\\
\textbf{Benchmarking.} The improvement brought about by optimal pulse profiling is characterized from gate-level, calibration-level, device-level, and application-level.

\subsection{Multi-derivative Drag}
\label{multiderivativedrag}

Ideally, the CR interaction leads to a rotation in the target qubit status and leaves the control qubit unaltered. However, the off-resonant microwave drive may excite the control qubits during the operation, especially when the frequency detuning between the qubits is smaller than their anharmonicities. To overcome
the unwanted excitations, a term proportional to the derivative of the drive pulse is introduced, the DRAG pulse, given by $\Omega - ia\frac{\dot{\Omega}}{\Delta}$.

For the off-resonant CR drive, all the transitions among the three states ($\ket{0}$, $\ket{1}$ and $\ket{2}$) of the control transmon need to be avoided simultaneously.
Therefore, multi-derivative DRAG was applied to suppress the undesired transitions by recursively applying the DRAG correction targeting three dominant transitions~\cite{Li2024Experimental,Motzoi2013Improving}
\begin{equation}
  \Omega_{\text{CR}}^{\text{P}} = \mathcal{F}_{\Delta_{21}}^{(1)} \circ \mathcal{F}_{\Delta_{10}}^{(1)} \circ \mathcal{F}_{\Delta_{20}}^{(2)}(\Omega)
\end{equation}
where $\Omega = \mathcal{F}_{\Delta}^{(n)}(\tilde{\Omega}) := \left( \tilde{\Omega}^n - i \frac{\text{d} \tilde{\Omega}^n}{\text{d}t} \frac{1}{\Delta} \right)^{\frac{1}{n}}$ and $\Delta_{jk}$ denotes the energy difference between state $\ket{j}$ and $\ket{k}$. The circle operator applies the pulse shape transformation $\mathcal{F}$ on top of the last pulse output.
The initial input pulse shape needs to be chosen so that the final pulse shape still starts and ends at zero~\cite{Li2024Experimental}.

Another major portion of the CR error derivatives from other dynamical operators in the two-qubit subspace that do not commute with the ideal dynamics ZX.
For instance, the unwanted terms in Equation ~\ref{eq:two-qubit CR ham} can be measured by Hamiltonian tomography and calibrated away, as in the standard CR pulse~\cite{sheldon2016procedure}.
The remaining terms, such as the $ZZ$ and $IZ$ error~\cite{Magesan2020Effective}, are also removed by DRAG corrections on the target qubit and adjusting the detuning~\cite{Li2024Experimental}.

Despite the improvement in gate quality, the multi-derivative pulse introduces additional calibration overhead. The calibration of the DRAG correction requires more tomography experiments and hence longer calibration time. The waveform generated under the multi-derivative DRAG tends to be more complicated, therefore challenging the hardware configuration and resource optimization for current quantum hardware. During calibration experiments on real quantum hardware, a preprocessing error is sometimes detected due to the overwhelmingly complicated waveform. More importantly, through numerical simulations, previous work has proved that multi-derivative DRAG waveform achieves the most significant improvement in a certain range of qubit-qubit detuning, e.g. as shown in Figure~\ref{fig:effective_range}. Consequently, applying multi-derivative DRAG may be unnecessary for qubits with frequency detunings outside of this interval.
\begin{figure}[t]
    \centering
    \includegraphics[width=\linewidth]{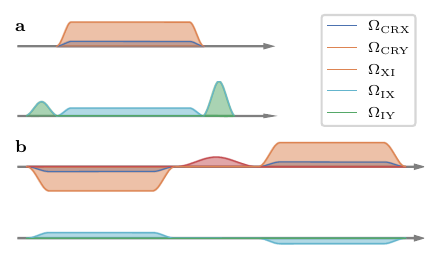}
    \includegraphics[width=\linewidth]{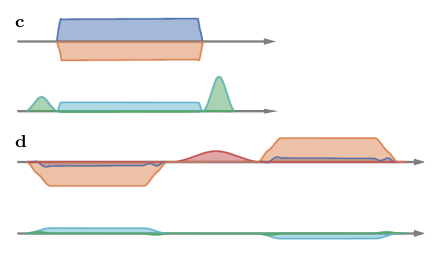}
    \caption{Examples of the CR pulse: the default pulse shape for direct (a) and echoed (b) CR as well as the multi-derivative direct CR (c) and echoed (d). The pulse amplitudes are rescaled for visualization purposes.
    }
    \label{fig:DRAG-CR-pulse-example}
\end{figure}

\subsection{Calibration Policy Design}
\label{calibrationpolicydesign}
The proposed approach involves three pulse waveforms, and a straightforward solution would be to conduct calibration for all three waveforms across every qubit pair. 
However, this method would be resource-intensive and time-consuming. 
Due to hardware limitations and external disturbances, modern quantum hardware experiences considerable error drift; specifically, qubit pairs typically reach an error level \textbf{5}$\times$ their initial value within approximately \textbf{20 hours}. 
To maximize fidelity while minimizing calibration time, an efficient policy should be developed to identify and calibrate the optimal waveform for each qubit pair.

\begin{figure}
    \centering
    \includegraphics[width=\linewidth]{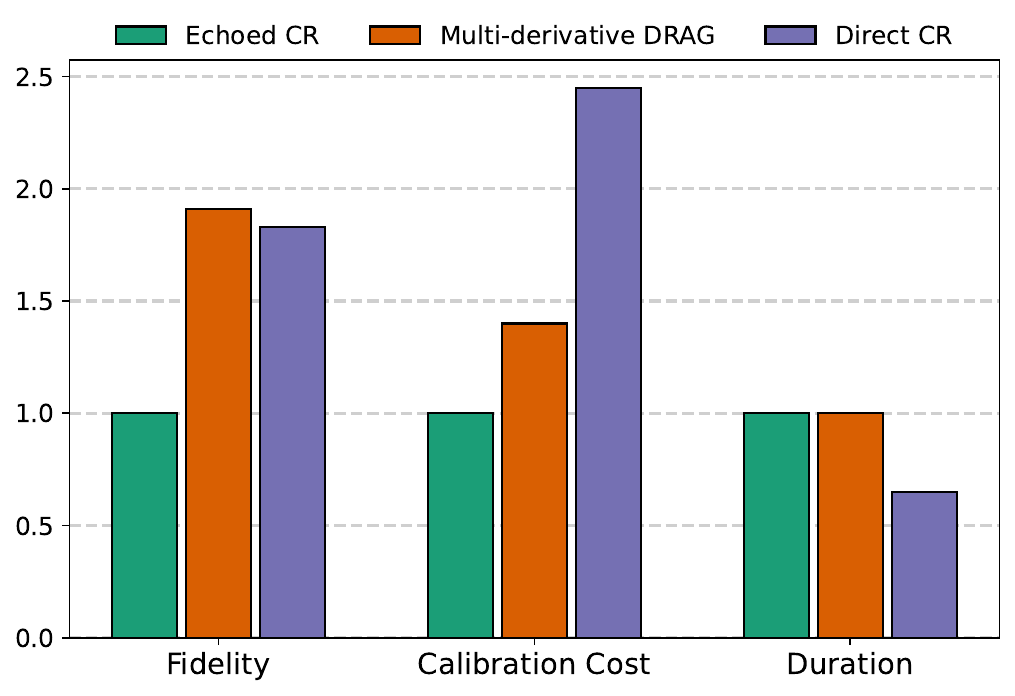}
    \caption{Fidelity, calibration cost, and duration trade-off among various ECR gate waveforms.}
    \label{fig:trade-off}
\end{figure}

\subsubsection{Waveform Candidates and Analysis} 
Based on previous discussions, any ECR gate could have multiple implementations, including the echoed CR waveform, the Multi-derivative DRAG waveform, and also the Direct CR waveform. Though the Direct CR waveform is calibrated towards the goal of realizing a CX operation, it could also be utilized to implement an ECR gates' waveform with an SX waveform and an X waveform on the target qubit. The cross-resonance part in the Direct CR can also be designed as containing multi-derivative parts. Therefore, the waveform candidates to realize an ECR gate are enlarged to four, as shown in Figure~\ref{fig:DRAG-CR-pulse-example}. However, calibrating the original direct CR waveform on real quantum hardware has been found to be exceedingly resource-intensive. As a result, in the following discussions, the Direct CR is implemented with multi-derivative parts.

Though various waveforms can be utilized to realize the same gate, a trade-off between gate fidelity, calibration cost, and duration exists among various waveforms. Multi-derivative DRAG  and Direct-CR waveforms share a potential to achieve a higher fidelity than the echoed CR, while the prior introduces a 1.4 times increase in calibration cost, and the latter one requires a calibration cost as high as 2.8 times (Figure~\ref{fig:trade-off}). However, the short duration of the Direct CR waveform is particularly critical for qubits with limited decoherence times, as it helps mitigate coherence loss during operations. Therefore, an important problem exists as profiling each qubit pair with its own optimal pulse waveform.
\begin{figure}[t]
    \centering
    \includegraphics[width=1\linewidth]{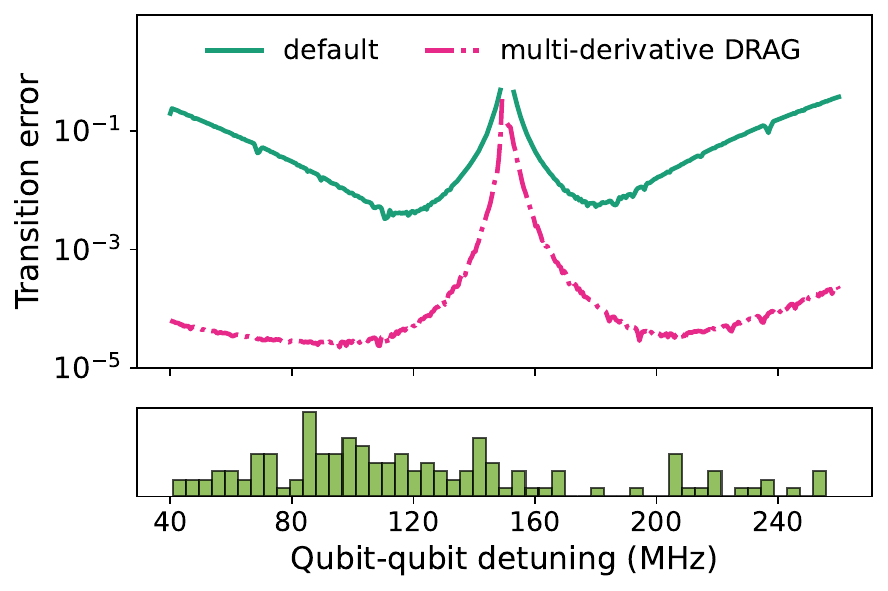}
    \caption{Top: Relationship between remaining transition errors and the frequency detuning for multi-derivative DRAG pulse. 
    The large error in the middle is located at half of the control qubit's anharmonicity, induced by a two-photon transition. 
    Bottom: Distribution of the qubit-qubit detuning on ibm\_brisbane. A few outliers are removed for clarity.}
    \label{fig:effective_range}
\end{figure}

\subsubsection{Brute-force Clustering.}

CR pulses use microwave to trigger Rabi oscillation on the target qubits depending on the state of the control qubit. 
However, the effect of Rabi oscillation and corresponding errors depend heavily on the physical properties between qubits. 
For example, a strong drive signal can often lead to coherent errors induced by non-adiabatic dynamics. 
To capture the interactive Hamiltonian between adjacent qubits, several critical metrics are considered, including the detuning between qubits' frequency, the anharmonicity, and the coupling strength between the qubits.
As shown in Figure \ref{fig:effective_range}, the transition errors on the control qubit depend strongly on the qubits detuning and the anharmonicity.
Apart from the control qubit error, the dominant coefficients in the effective Hamiltonian in Equation~\ref{eq:two-qubit CR ham} are also decided by the frequency detuning and coupling strengths, given by
\begin{equation}
  H_{CR} = -\frac{\Delta_{12}}{2} \hat{Z}\hat{I} + 
  \frac{\Omega (t)}{2}(\hat{I}\hat{X} - \frac{J}{2 \Delta_{12}}\hat{Z}\hat{X})
\end{equation}
where $\Delta_{12}$ is the frequency detuning between two qubits and $\Omega(t)$ is the cross resonance drive strength.
\begin{figure}[t]
    \centering
    \begin{subfigure}[t]{0.49\linewidth}
        \centering
        \includegraphics[width=\textwidth]{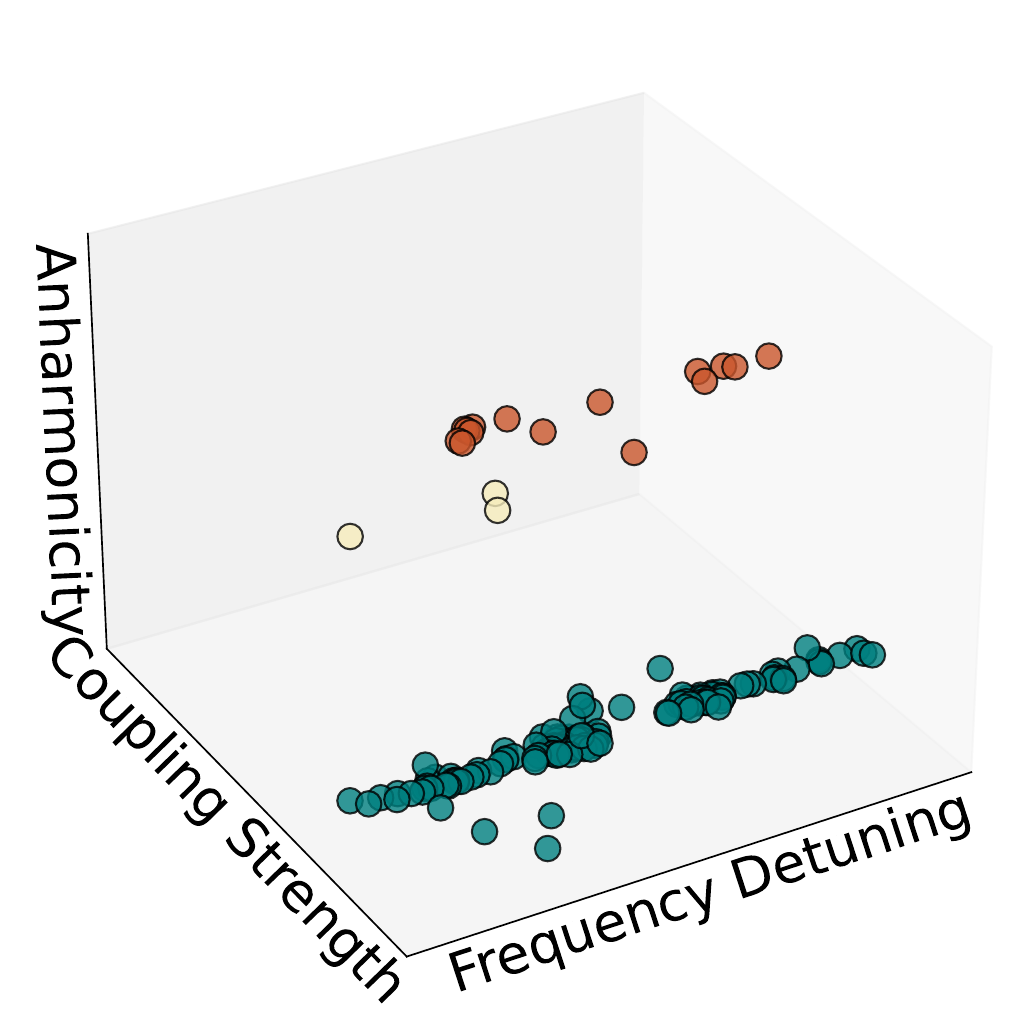} 
        \caption{ibm\_rensselaer, $n$=3}
        \label{fig:ren_3}
    \end{subfigure}
    \begin{subfigure}[t]{0.49\linewidth}
        \centering
        \includegraphics[width=\textwidth]{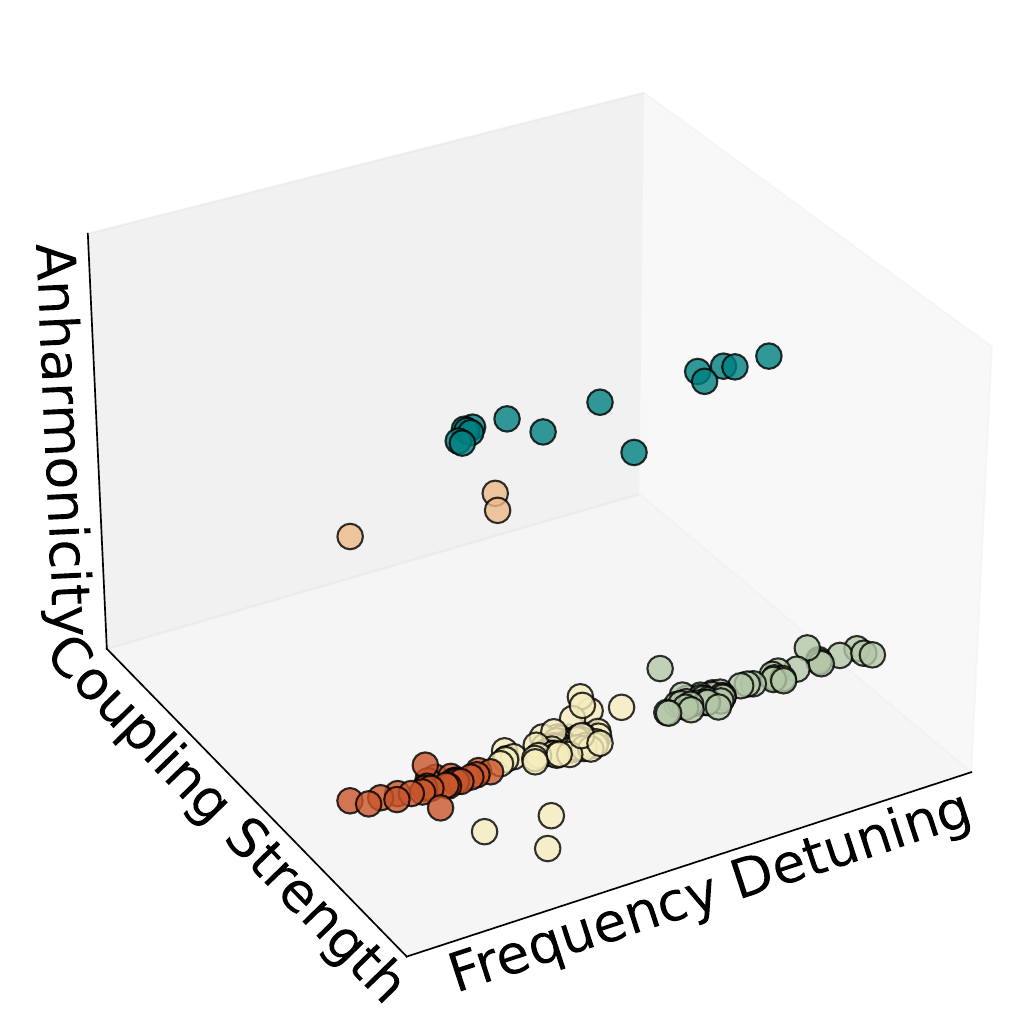} 
        \caption{ibm\_rensselaer, $n$=5}
        \label{fig:ren_5}
    \end{subfigure}
    \begin{subfigure}[t]{0.49\linewidth}
        \centering
        \includegraphics[width=\textwidth]{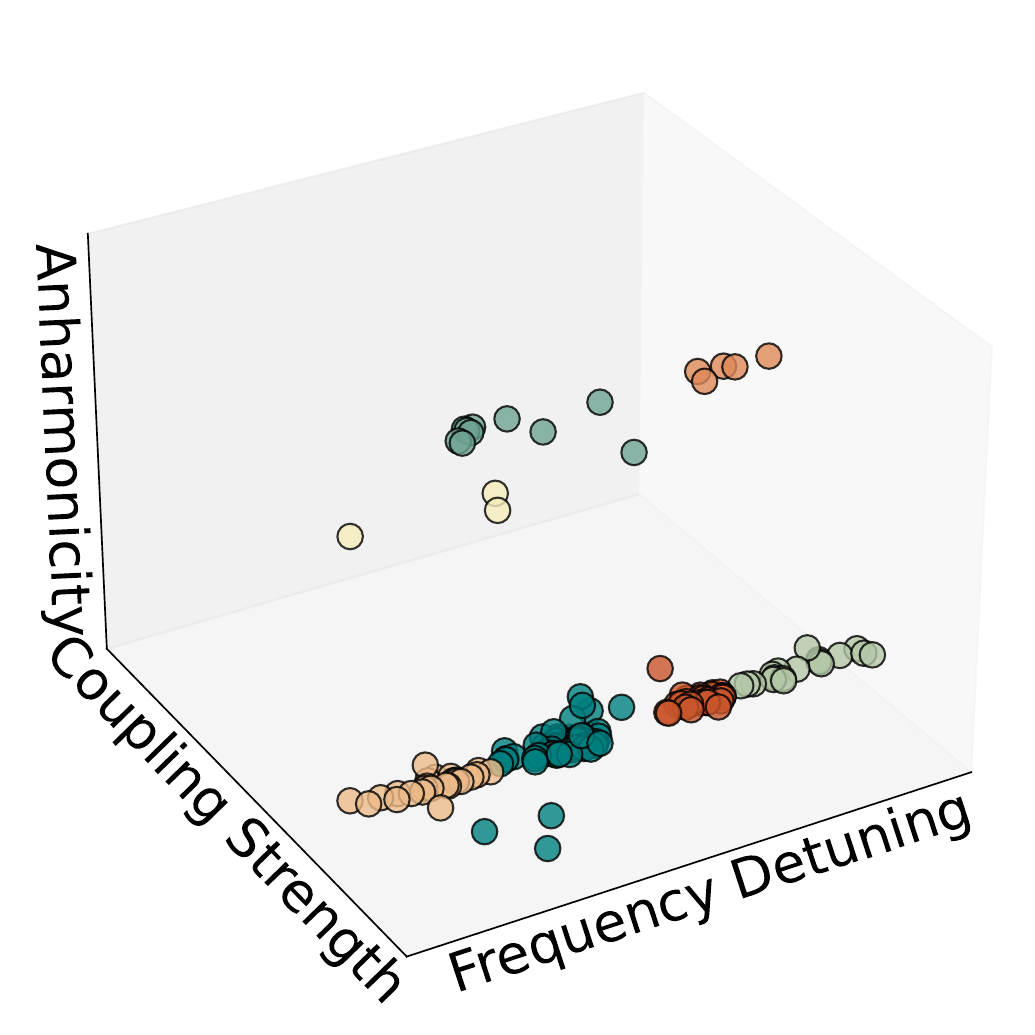} 
        \caption{ibm\_rensselaer, $n$=7}
        \label{fig:ren_7}
    \end{subfigure}
    \begin{subfigure}[t]{0.49\linewidth}
        \centering
        \includegraphics[width=\textwidth]{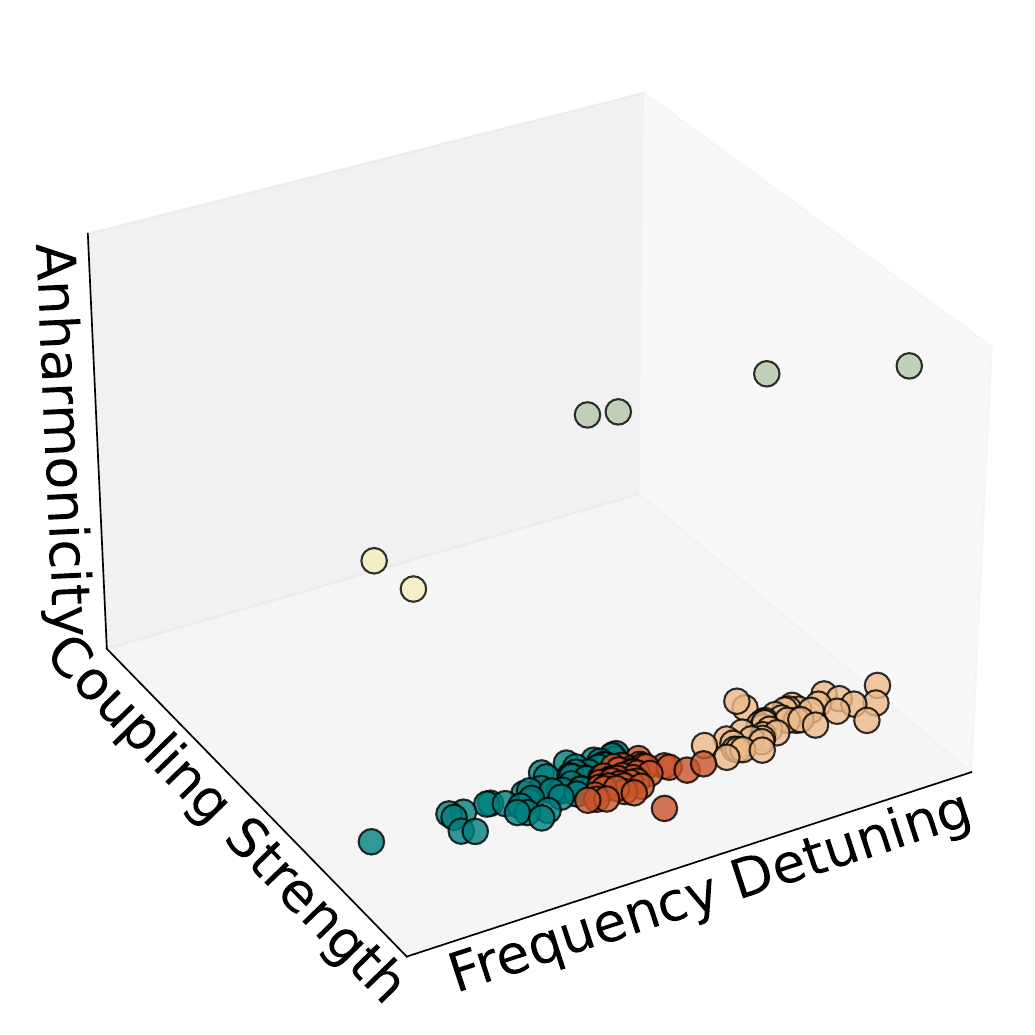} 
        \caption{ibm\_nazca, $n$=5}
        \label{fig:naz_5}
    \end{subfigure}
    \begin{subfigure}[t]{0.49\linewidth}
        \centering
        \includegraphics[width=\textwidth]{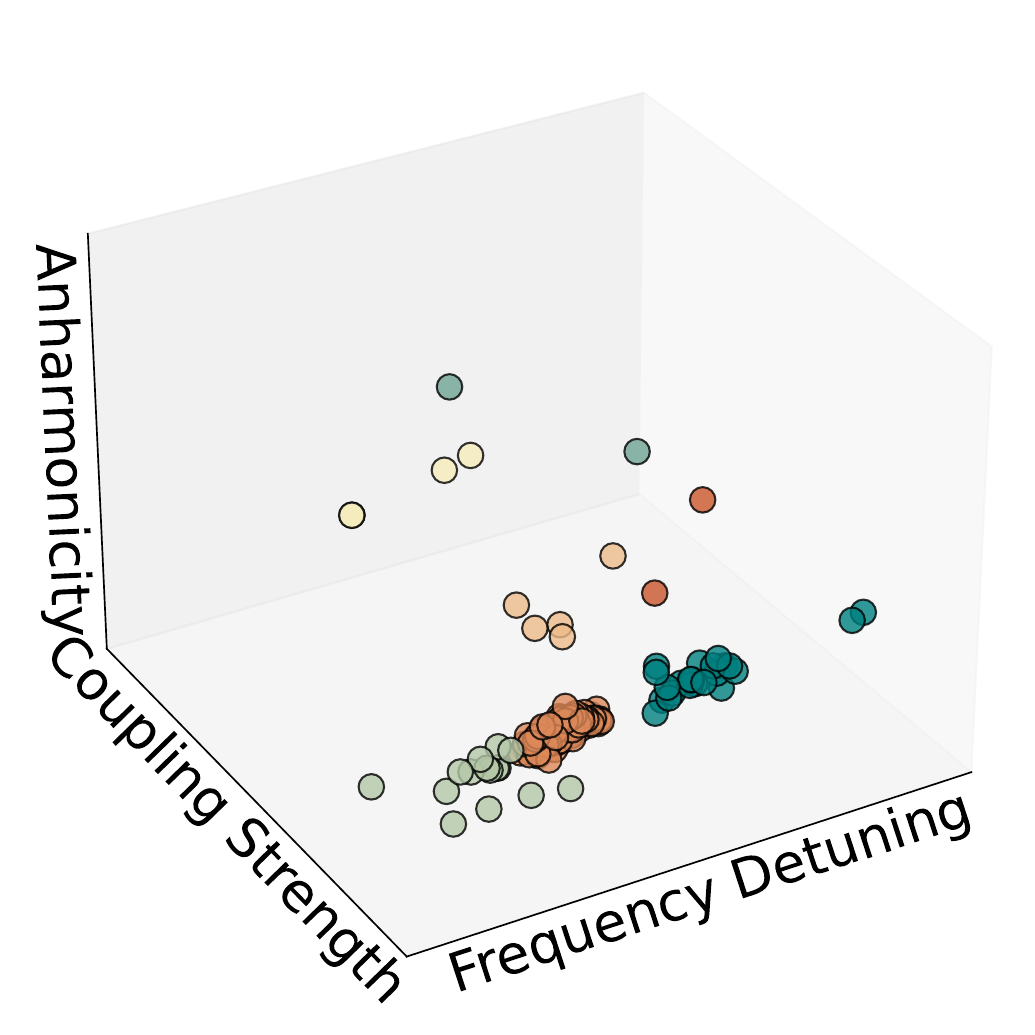} 
        \caption{ibm\_sherbrooke, $n$=7}
        \label{fig:she_7}
    \end{subfigure}
    \begin{subfigure}[t]{0.49\linewidth}
        \centering
        \includegraphics[width=\textwidth]{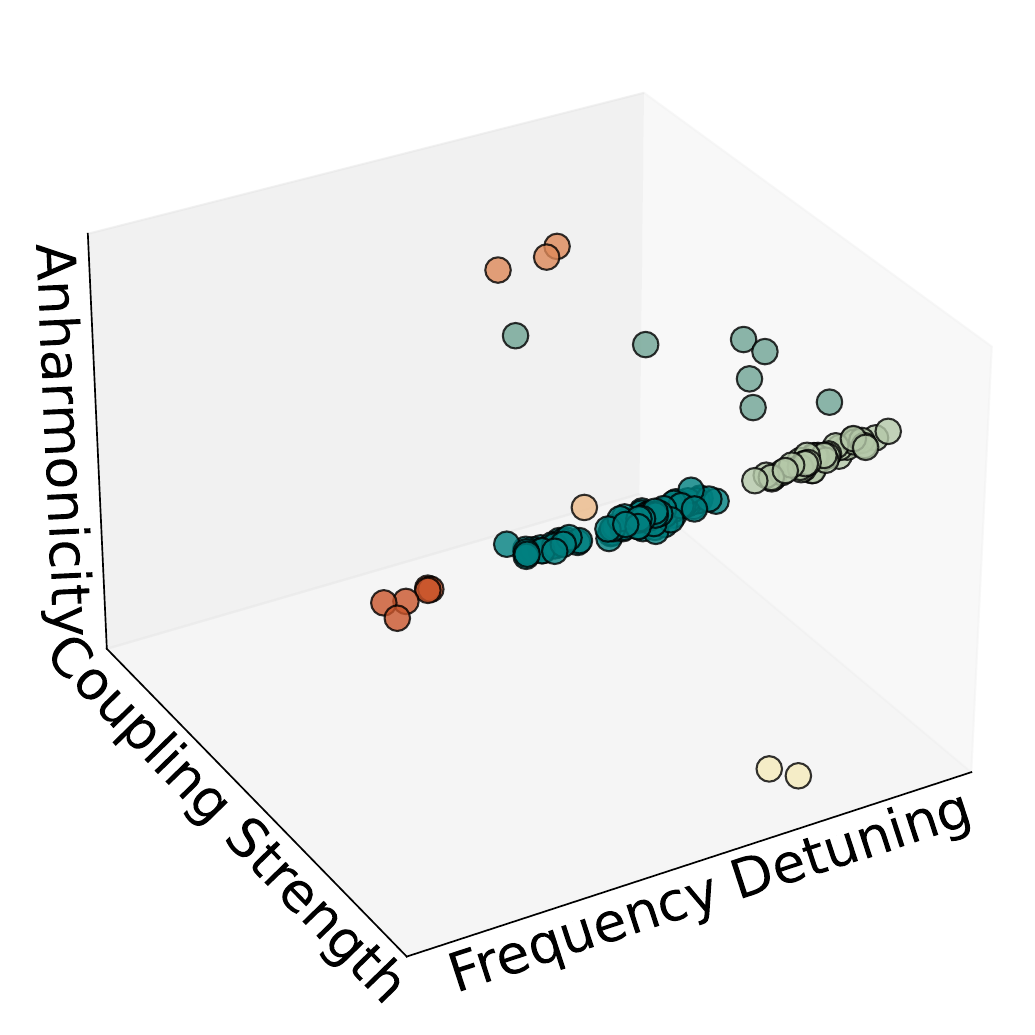} 
        \caption{ibm\_brisbane, $n$=7}
        \label{fig:bri_7}
    \end{subfigure} 
    \caption{Clustering results for various quantum devices (ibm\_rensselaer, ibm\_nazca, ibm\_sherbrooke, and ibm\_brisbane) based on frequency detuning, coupling strength, and anharmonicity for various clustering size $n$.}
    \label{fig:cluster}
\end{figure}

The frequency detuning, coupling strength, and anharmonicity of target qubits are prioritized as major features when profiling the optimal CR pulse. 
Each qubit pair is projected into a multi-dimensional vector, which captures the physical properties that are critical to high-fidelity control. 
Those vectors are clustered with the Birch algorithm into multiple groups, and partition qubit pairs into sets with similar characteristics. 
Then one representative qubit pair is selected from each group and three calibration processes are performed on the representatives.
The election of the representative allows us to identify the control waveform with optimal fidelity that can be generalized to all pairs within the cluster. 
To balance calibration efficiency and fidelity, the cluster size is selected to be three, five, and seven.
We note that fewer groups result in a faster calibration process, and more groups would lead to a higher accuracy. 
We implement this approach on multiple real quantum devices, the results are illustrated in Figure~\ref{fig:cluster}.
The clustering-based approach optimally distributes calibration resources ensures an effective trade-off between computational efficiency and calibration accuracy across qubit pairs.

\subsubsection{Topology-oriented Representative}
The Brute-force clustering method manages to strike a balance between calibration cost and the accuracy for waveform profiling. However, the performance of Brute-force clustering depends heavily on the clustering size, which is a hyperparameter predefined, and it is uncertain how many clusters are required to meet the sweet spot of the method for various quantum computer systems. 

\begin{figure}
    \centering
    \includegraphics[width=0.49\linewidth]{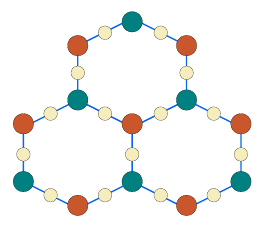}
    \caption{Repetitive pattern of the heavy-hex lattice, colors indicate the pattern of distinct frequencies for control qubits (pale yellow) and two sets of qubits (orange and deep teal).}
    \label{fig:hex_sim}
\end{figure}

Recently, the topology of all active IBM quantum devices is based on the heavy-hex lattice, where each unit cell of lattice consists of a hexagonal arrangement of qubits, with an additional qubit on each edge. For CR operations, the resonance frequency executed on the target qubit needs to be precisely off-resonant with neighboring qubit transition frequencies to prevent unwanted interactions, also known as frequency collisions. Compared with the traditional square lattice model, the reduced connectivity of heavy-hex topology brings about fewer frequency constraints, providing a better chance at achieving higher fidelity CR control gates.
Previous research has highlighted a recurring pattern among unit cells in the heavy-hex lattice topology~\cite{smith2022scaling, noauthor_ibm_nodate}, where qubits occupying analogous positions across different unit cells exhibit similar physical characteristics. This regularity is a result of the uniform structure of the heavy-hex architecture, where each unit cell maintains a consistent arrangement, leading to shared properties such as frequency detuning, coupling strength, and anharmonicity among qubits in equivalent positions. Figure~\ref{fig:hex_sim} depicts such a similarity where colors indicate the pattern of  distinct frequencies for control and target. Leveraging this structural regularity, each edge in the heavy-hex topology can be classified according to its relative position within the unit cell. By categorizing edges in this way, the calibration complexity is effectively reduced. For each distinct position within the local unit cell, we select a representative edge to undergo the complete three-stage calibration process. This allows us to identify the waveform that achieves optimal fidelity, which can then be generalized to all edges in equivalent positions across the lattice. This approach significantly streamlines the calibration process by minimizing redundant calibrations, making it possible to apply a high-fidelity waveform to all similar edges with greater efficiency. Compared with the Brute-force clustering, such a method limits the cluster size to 12 and the classification policy not only includes evident qubit physical parameters but also considers the hidden constraints generated by hardware topology. 

\subsubsection{Hardware-oriented Policy}
Previous solutions have generated waveform profiling for all qubit pairs towards the highest fidelity while limiting a reasonable calibration time. Another critical factor, system knowledge, has been underscored by previous research as essential for optimizing performance on real quantum hardware~\cite{klimov2020snake}, though it was not addressed in previous discussions. Typical elements of system knowledge encompass the sequential order of calibrations required to initialize and optimize a qubit, as well as targeted optimizations tailored to specific qubit pairs. The calibration sequence involves channel frequency calibrations and single-qubit calibrations, which are out of the scope of this paper. On the other hand, targeted optimizations could contribute majorly to a more efficient and accurate waveform design. In the hardware-oriented policy, system knowledge of the relationship between frequency detuning and waveform is selected as a major constraint when profiling pulse waveforms. Through numerical simulations, it’s exploited that for qubit pairs that reside outside a specific frequency range, calibrating multi-derivative waveform takes much longer time than echoed CR pulse and fails to eliminate error terms to an ideal extent (0.015MHz). Therefore, for such qubit pairs, the echoed CR waveform should be directly adopted to increase the efficiency of the calibration process and improve the calibration accuracy.

\begin{figure}[t]
    \centering
    \begin{subfigure}[t]{0.9\linewidth}
        \centering
        \includegraphics[width=\textwidth]{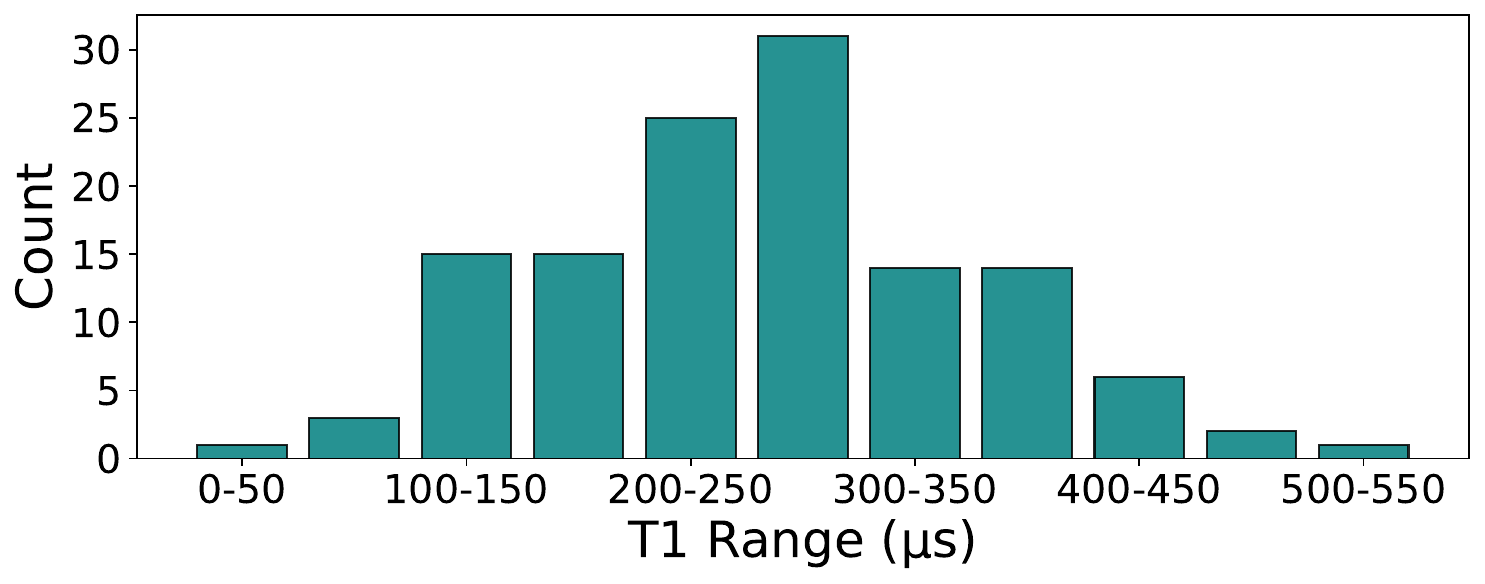} 
        \caption{T1 distribution for ibm\_rensselaer}
        \label{fig:ren_t1}
    \end{subfigure}
    \begin{subfigure}[t]{0.9\linewidth}
        \centering
        \includegraphics[width=\textwidth]{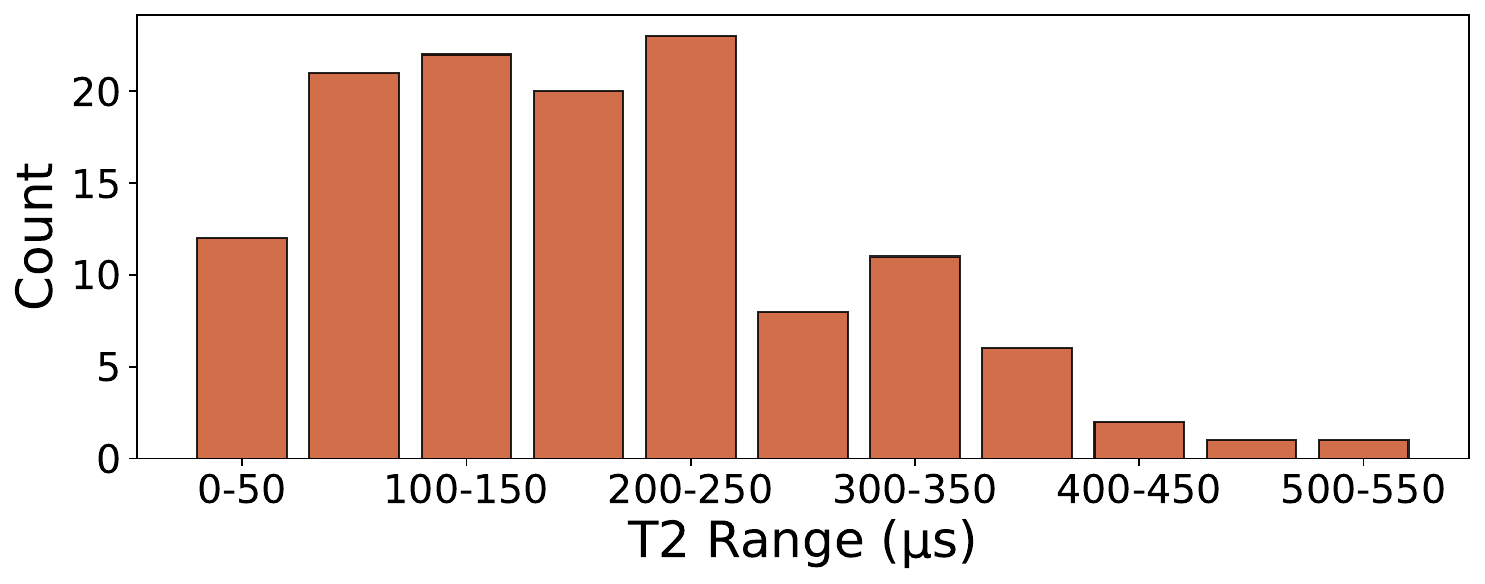} 
        \caption{T2 distribution for ibm\_rensselaer}
        \label{fig:ren_t2}
    \end{subfigure}
    \caption{Longitudinal relaxation (T1) and transverse relaxation (T2) time distribution for ibm\_rensselaer, indicating an non-negligible number of qubit pairs with T1 or T2 below 150$\mu s$.}
    \label{fig:ren_t1t2}
\end{figure}

Another crucial metric that is missed from previous discussions is the decoherence limit of current quantum computer systems, which exists as an important limit for the duration of quantum programs. The decoherence limit includes the longitudinal relaxation (T1) and the transverse relaxation (T2), which describes how long the excited state and the superposition state of the qubits can be preserved, respectively. As we can observe from Figure \ref{fig:ren_t1t2}, for current quantum hardware, the median of T1 and T2 are 269 $\mu s$ and 172 $\mu s$, respectively. A typical pulse of ECR gate takes approximately 665 ns to execute. For an average-quality qubit pair, this implies that the theoretical upper bound on the number of sequential ECR gates that could produce meaningful results is about 258. However, the actual scenario is more severe as the general performance of a quantum program is limited by qubit pairs with the worst performance. For contemporary hardware systems with 127 qubits, the minimal T2 could be only 20$\mu s$, and around 20 qubit pairs bear a decoherence limit of less than 60$\mu s$. For those qubit pairs, the effectiveness of various waveform strategies is constrained not only by the achievable fidelity but also by the duration of the pulses. Specifically, while an echoed CR pulse may achieve the highest fidelity, its longer duration often renders it less practical than a direct CR pulse. The latter, typically requiring around 60\% to 80\% the duration of the former, tends to be the preferred choice in scenarios where minimizing pulse duration is critical to overall performance.

\subsection{Graph traverse and parallelization}
\label{graphtraverse}
\begin{figure}
    \centering
    \includegraphics[width=\linewidth]{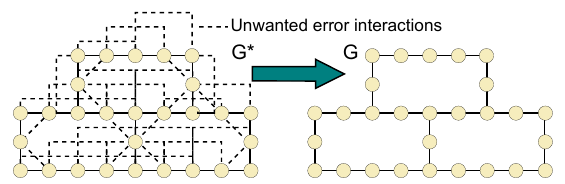}
    \caption{Calibration target: remove unwanted interactions from a connected graph to reach its subgraph, where all edges are calibrated with optimal parameters.}
    \label{fig:calib_subgraph}
\end{figure}

After individually profiling each qubit pair with its optimal CR pulse waveform, a significant challenge remains: calibrating all qubit pairs within a specific time frame without stalling the quantum machine for too much time. A coupling graph with $N$ nodes and $E$ edges is seen as an undirected graph $G$. The scheduling problem could be redefined as learning the optimal pulse parameter for some subset of edges in the coupling graph while limiting unwanted interactions outside the subset, which is referred to as $G^* \in G$. Calibration is accomplished by traversing $G$ while calibrating parameters associated with each edge in the target $G^*$ as shown in Figure \ref{fig:calib_subgraph}.

\begin{figure}
    \centering
    \includegraphics[width=\linewidth]{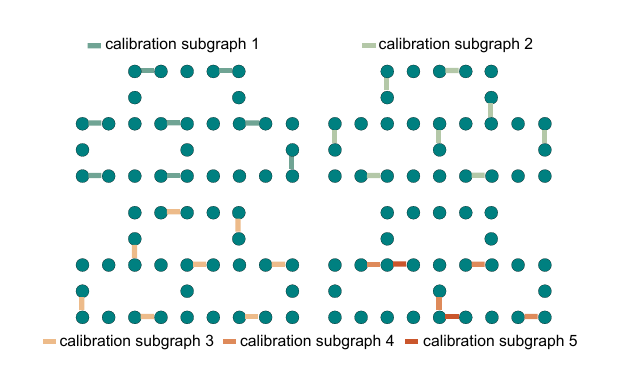}
    \caption{Dividing a heavy-hex coupling graph into five calibration subgraphs for parallel graph traversal.}
    \label{fig:calib_subgraph2}
\end{figure}

As provided in Figure \ref{fig:calib_subgraph2}, the spacetime structure of the calibration process is defined as calibration subgraphs. Each subgraph represents the part of the graph which is simultaneously calibrated. The separation of calibration subgraphs depends on the elements that could interfere with each during the calibration process. When calibrating CR pulses, it can be seen as executing ECR gates. Therefore, the calibration subgraph could be decided as the maximum number of ECR gates that could execute at the same circuit layer. In general, the total graph $G$ is traversed after the calibration, and at each step, all edges at the subgraph are calibrated. To build the calibration subgraphs, all single paths with a length of one are selected from the coupling map. Then they are separated with the requirement of a minimum distance of two for all edges in a single calibration subgraph. For the heavy-hex structure with 127 qubits, the number of the calibration subgraph could be limited to five, with a maximum number of 38 qubit pairs calibrating simultaneously.

\begin{figure}
    \centering
    \begin{subfigure}[t]{\linewidth}
        \centering
        \includegraphics[width=\textwidth]{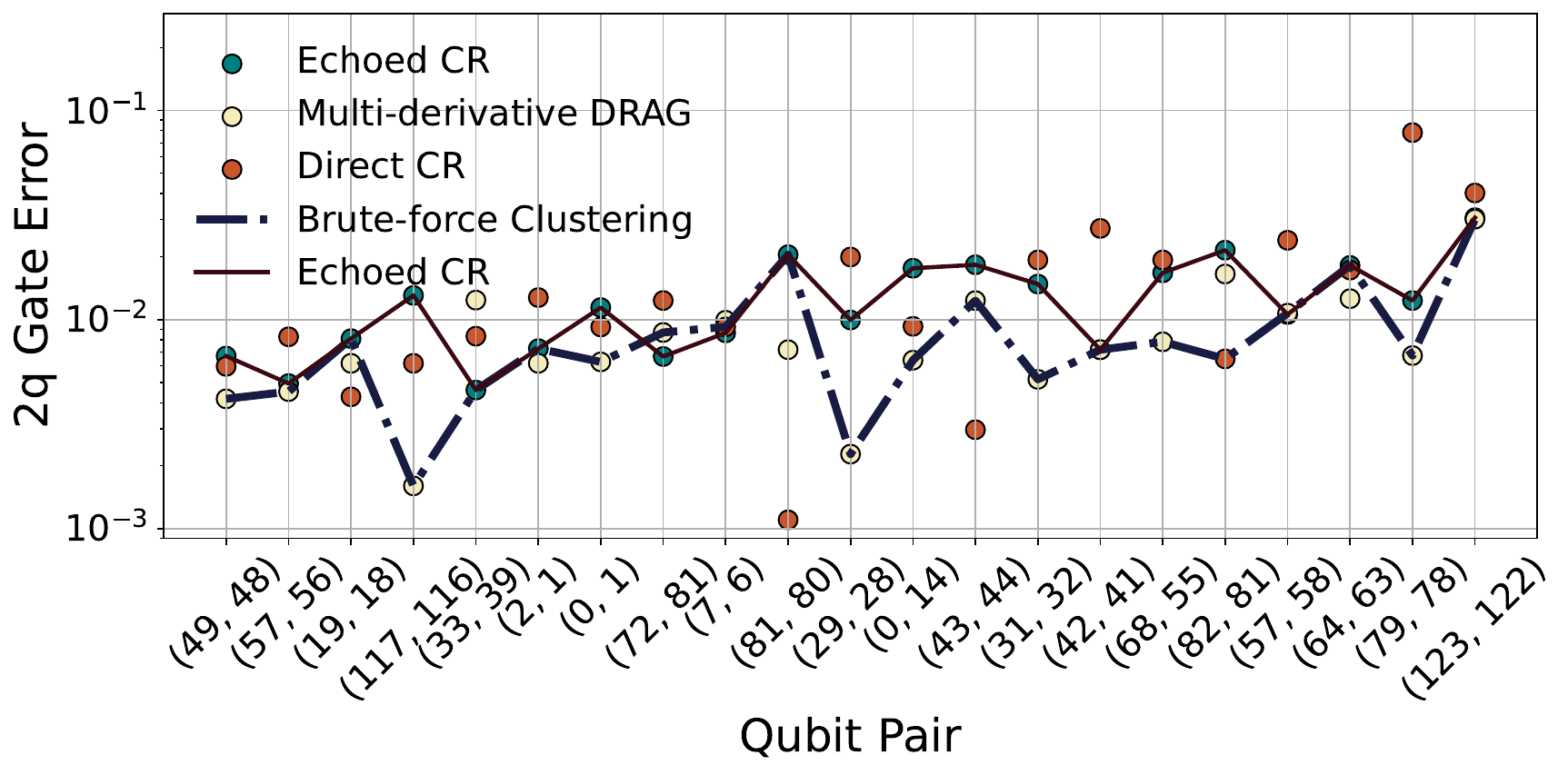} 
        \caption{Performance of Brute-force clustering with clustering size of 7. Echoed CR pulse is the default pulse waveform by IBM.}
        \label{fig:policy1}
    \end{subfigure}
    \begin{subfigure}[t]{\linewidth}
        \centering
        \includegraphics[width=\textwidth]{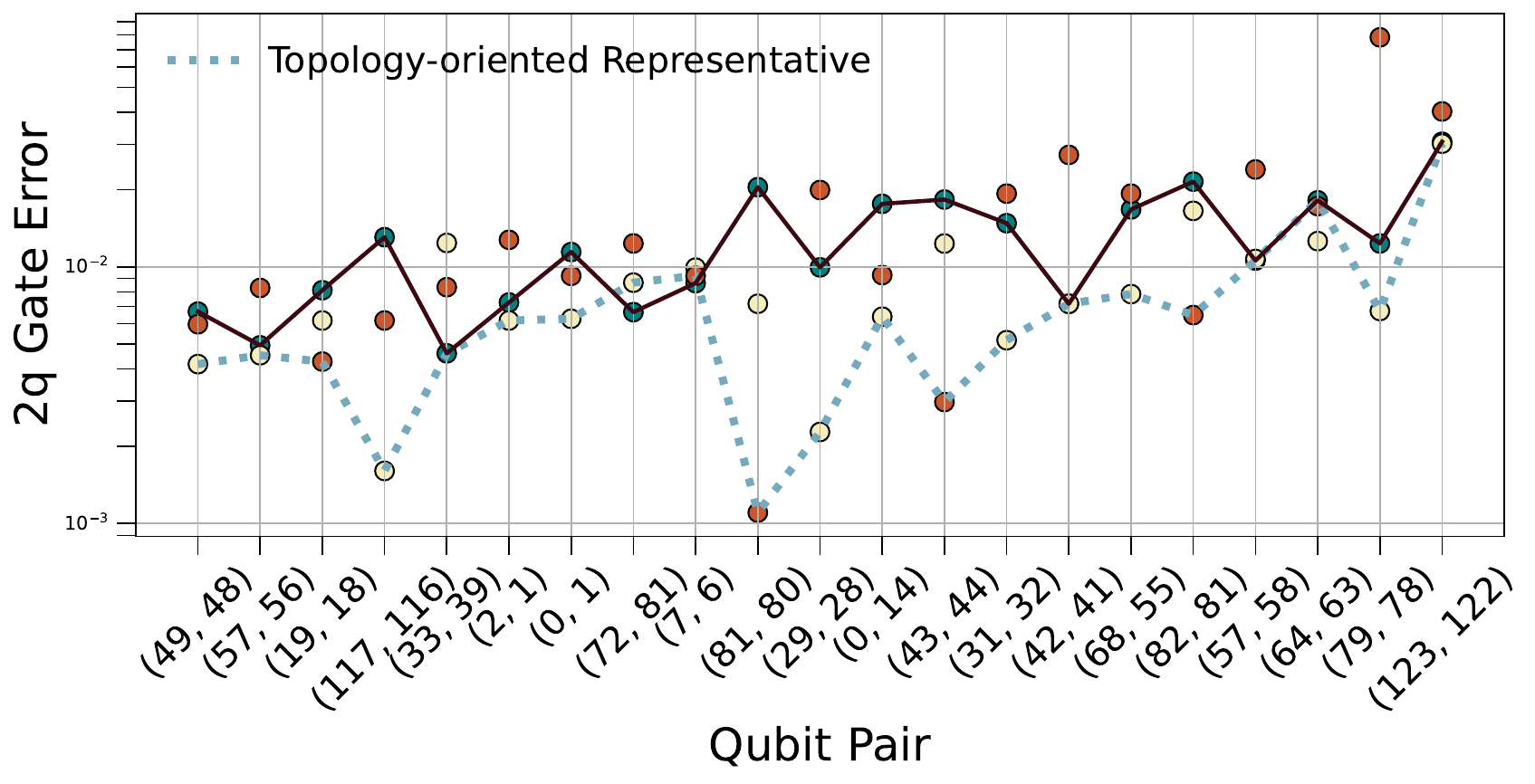} 
        \caption{Performance of Topology-oriented Representative.}
        \label{fig:policy2}
    \end{subfigure}
    \begin{subfigure}[t]{\linewidth}
        \centering
        \includegraphics[width=\textwidth]{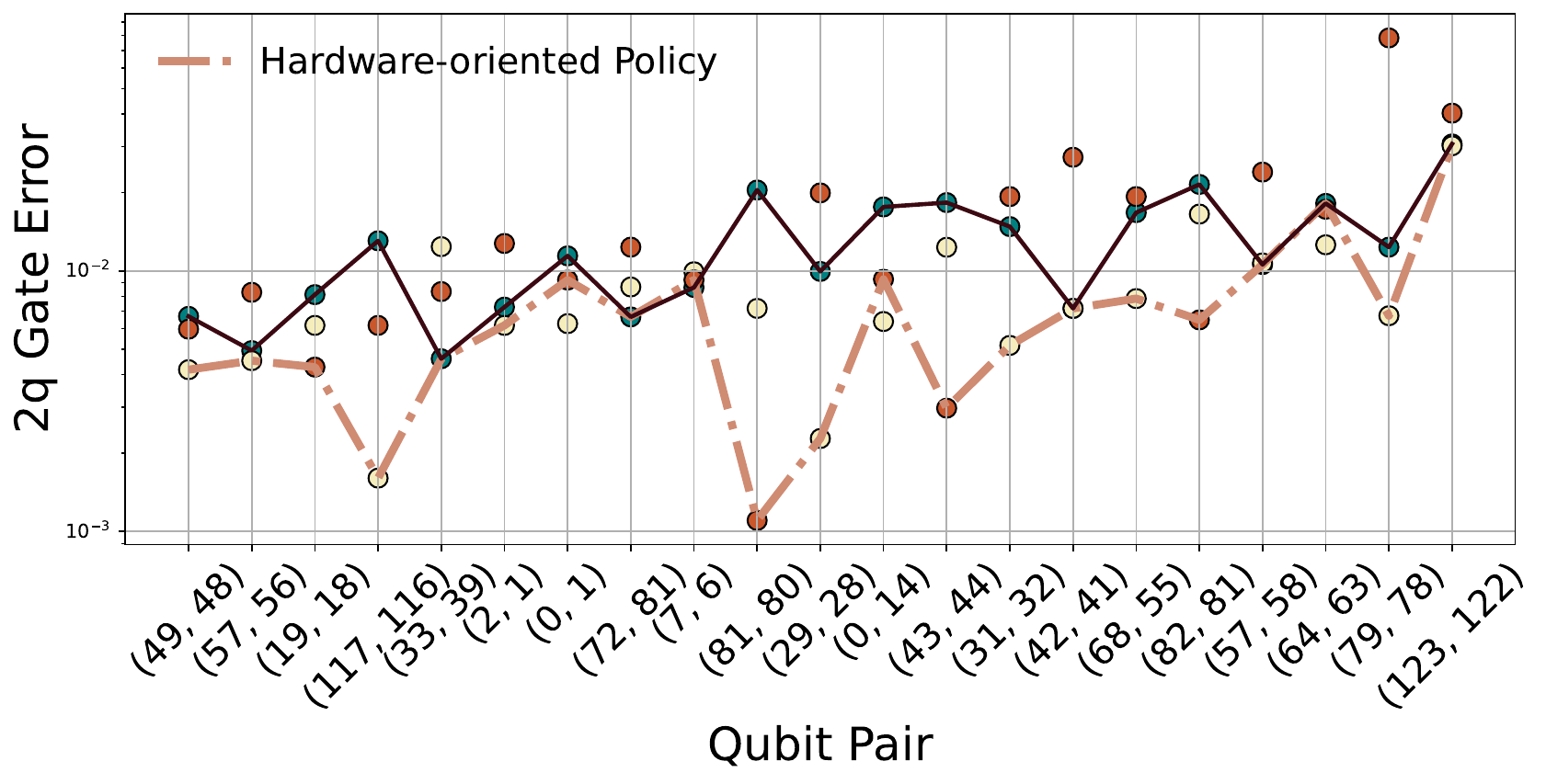}
        \caption{Performance of Hardware-oriented Policy.}
        \label{fig:policy3}
    \end{subfigure}
    \begin{subfigure}[t]{\linewidth}
        \centering
        \includegraphics[width=\textwidth]{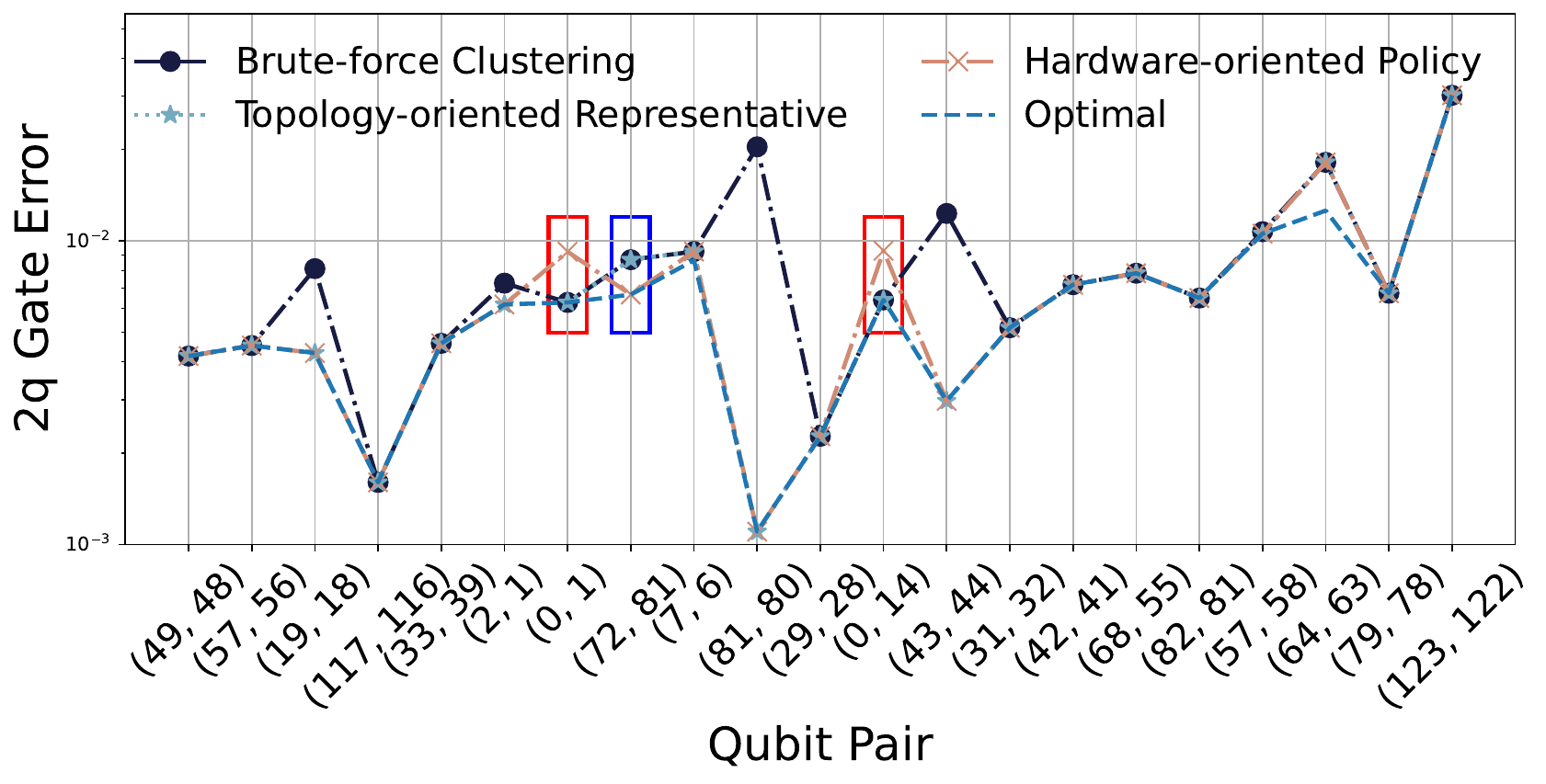}
        \caption{Horizontal comparison of the three profiling policies.}
        \label{fig:hor_comp}
    \end{subfigure}
    \caption{Performance benchmarking of various waveform profiling policies on randomly selected qubit pairs. }
    \label{fig:bench}
    
\end{figure}

\section{Evaluation}
\label{evaluation}

\subsection{Configuration}
\label{configuration}
All experiments are conducted on Eagle r3 quantum processors provided by IBM quantum service, including ibm\_rensselaer, ibm\_nazca, and ibm\_strasbourg. Eagle r3 processor is a family of superconducting-based quantum processing units. They share a configuration of 127 qubits and a heavy-hex coupling graph.

We implement Echoed CR pulses using symbolic functions in Qiskit. In contrast, multi-derivative DRAG pulses and Direct CR pulses are sent to the hardware as arrays of pulse amplitudes. Due to limitations in IBM quantum machines, we split multi-derivative DRAG pulses into two parts to avoid overly complicated custom pulse shapes. We use SciPy~\cite{virtanen2020scipy} to process Hamiltonian tomography results and optimize pulse parameters.
Before calibrating ECR gates, we focus on qubits whose single-qubit gate error rates were significantly higher than the device median. These calibrated single-qubit gates are later used in Hamiltonian tomography experiments. We initially set an error threshold of 0.015 MHz for all calibration experiments. If qubit pairs fail to meet this threshold after four calibration rounds, we increase the threshold to 0.3 MHz. Our results show that over 99\% of qubit pairs could limit error terms to 0.3 MHz within four calibration rounds.
To accurately measure the errors of ECR gates before and after calibration, we employ Interleaved Randomized Benchmarking (IRB) to assess waveform fidelity. Each IRB experiment uses sequence lengths of 1, 10, 20, 50, 100, 150, 250, and 400. We repeat each experiment five times for each pulse configuration to calculate the mean and standard deviation of the gate error. Therefore, the reported gate error represents the average error over the hours following calibration, which has included any drift in system parameters during that time.

\subsection{Gate-level Benchmarking}
\label{accuracyofpolicydecision}
In all policies for profiling optimal ECR gates' waveforms, the primary goal is to achieve maximum fidelity across all qubit pairs. To evaluate whether this can be uniformly accomplished, we performed a comprehensive calibration of three waveforms on each qubit pair, beyond the standard calibration according to each policy.

Benchmarking results, displayed in Figure~\ref{fig:bench}, include randomly selected qubit pairs with error terms restricted to 0.015 MHz and 0.03 MHz. For the 21 qubit pairs selected, the Brute-force Clustering method showed a significant increase in fidelity compared to the default echoed CR pulse provided by IBM. However, for about five qubit pairs, this method failed to select the waveform with the highest fidelity, as seen in Figure~\ref{fig:policy1}.
In contrast, the Topology-aware Representative method demonstrated higher accuracy in pulse profiling, with only two-qubit pairs not achieving the optimal waveform, as illustrated in Figure~\ref{fig:policy2}. When comparing all three profiling policies in Figure~\ref{fig:hor_comp}, the Hardware-oriented Policy achieved nearly the same fidelity as the Topology-aware Representative. For certain qubit pairs, it favored the Direct CR waveform due to its significantly reduced gate duration. For example, in the left blue box, the Hardware-oriented Policy selects the Direct CR waveform for qubit pair (0, 1) because qubit 1 has a $T_2$ time of only 82.99 $\mu s$, which is less than half the median value. In the right box, Direct CR is chosen for the qubit pair (0, 14) since it offers similar fidelity to the Multi-derivative DRAG waveform while maintaining a shorter duration. Meanwhile, as shown in the middle blue box, for qubit pair (72, 81) with frequency detuning outside the optimal range for the Multi-derivative DRAG waveform, the Hardware-oriented Policy automatically selects the echoed CR as the optimal pulse, resulting in a lower error rate.
\begin{figure}
    \centering
    \includegraphics[width=0.9\linewidth]{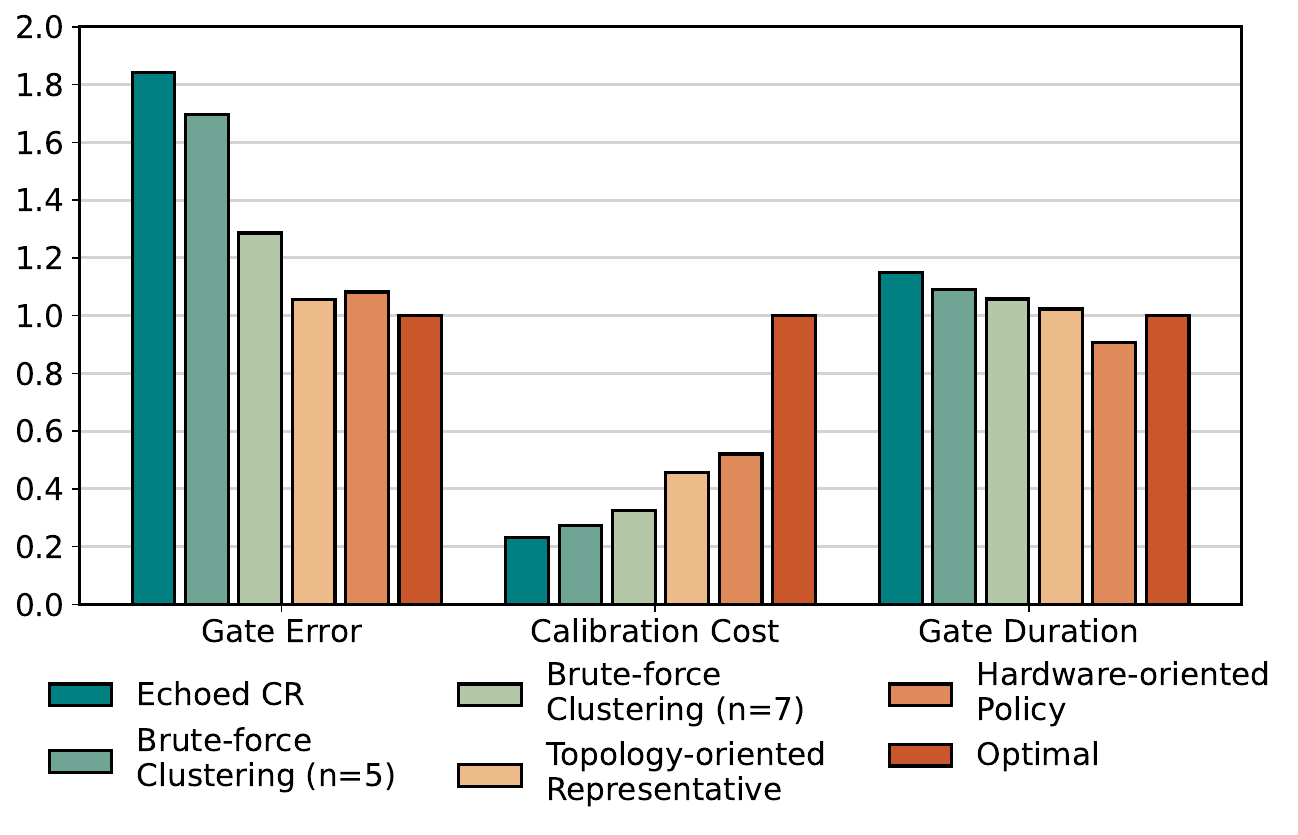}
    \caption{The sum of the gate error rate, calibration cost, and gate duration for all qubit pairs based on various profiling policies. All metrics are normalized to the optimal.}
    \label{fig:fidelity5}
\end{figure}

To evaluate the performance of different policies, we considered three main metrics for each qubit pair: the sum of error rates, calibration cost, and gate durations. Figure~\ref{fig:fidelity5} provides a detailed comparison of these metrics, with each normalized to its optimal value. The optimal calibration cost is calculated based on calibrating all three possible waveforms for every qubit pair.
The Topology-oriented Representative policy achieves a total error rate close to the optimal scenario, effectively selecting the waveform with the highest fidelity for nearly all qubit pairs. Meanwhile, its total calibration cost is less than half of what it would be if all three waveforms were calibrated for each pair. The Hardware-oriented Policy also attains a similar total error rate compared to the optimal solution. Furthermore, it results in a significantly shorter total gate duration than other profiling policies, which could enable the execution of longer gate sequences within the decoherence limit.

Combining waveform selection for specific qubit pairs and the generalized results for the whole quantum machine, our fine-tuned protocol can achieve almost the optimal error rate on all qubit pairs while limiting the total gate duration. 
After applying the whole calibration process on the real quantum machine, the medium of the two-qubit gate error rate is reduced to $4.4\times 10^{-3}$, representing a $1.84\times$ improvement compared to the default pulse configuration provided by IBM. 
For qubits with fabrication defects and shorter decoherence limits, the total pulse duration is reduced by a factor of 1.26. 
The proposed protocol could reduce the two-qubit gate error rate to the minimum at $1.3\times 10^{-3}$.
According to the real IBM quantum hardware results described in the~\cite{benito2024comparative}, this error rate is already below the two-qubit gate error rate threshold ($3\times 10^{-3}$).
It means that with our calibration method, the quantum error correction code has entered the region where errors are suppressed. 
\begin{figure}
    \centering
    \includegraphics[width=0.9\linewidth]{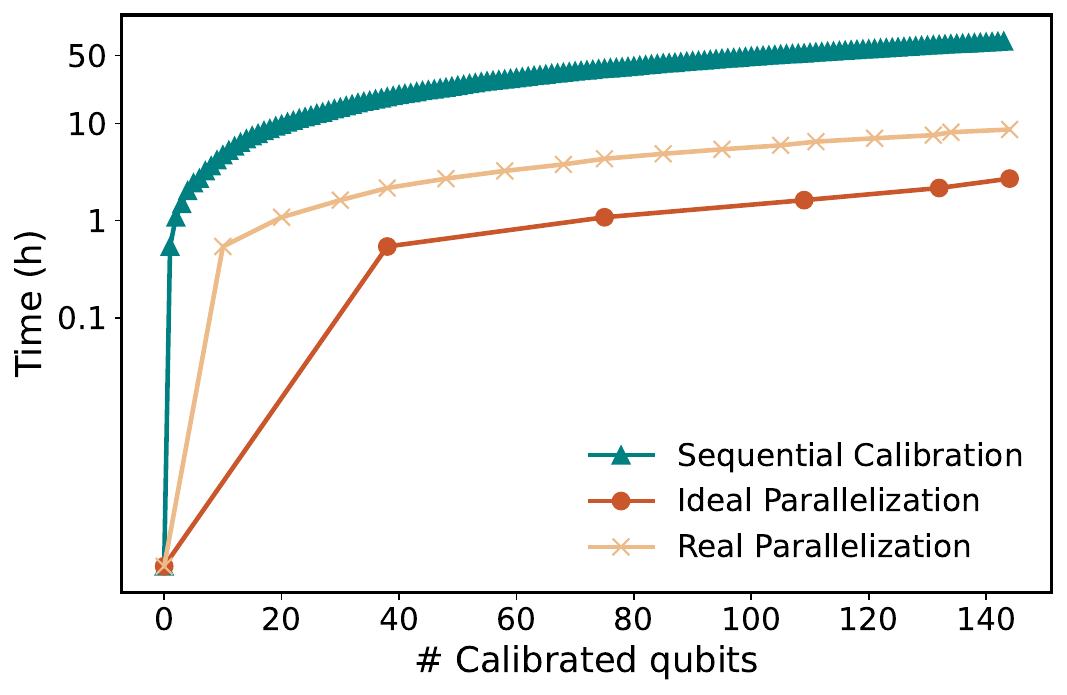}
    \caption{Calibration time comparison among sequential calibration, ideal parallel calibration, and limited parallel calibration due to hardware limitations.}
    \label{fig:calibrationtime}
\end{figure}

\begin{table}[t]
    \begin{tabular}{ccccc}
        \toprule
        \multirow{2}{*}{Quantum Device} &\multicolumn{2}{c}{\makecell{Quantum Volume}} 
        & \multicolumn{2}{c}{\makecell{Error Per\\Layered Gate}} \\
        
        & Default & Calibrated & Default & Calibrated \\
        \midrule
        ibm\_rensselaer & 128 &256 &3.08e-2 & 1.34e-2\\
        ibm\_nazca & 128 & 256 & 2.94e-2 & 1.48e-2\\
    \bottomrule
    \end{tabular}
    \caption{Device-level benchmarking results for ibm\_rensselaer and ibm\_nazca.}
    \label{tab:devicebenchmarking}
\end{table}

\begin{table*}[t]
\centering
    \begin{tabular}{ccccccccc}
        \toprule
        \multirow{2}{*}{Benchmark} &\multirow{2}{*}{\# Qubits} 
        & \multirow{2}{*}{\# ECR Gates} & \multirow{2}{*}{ECR Depth}
        & \multirow{2}{*}{Depth} 
        & \multicolumn{2}{c}{\makecell{Error Rate}} 
        & \multicolumn{2}{c}{\makecell{Fidelity}} \\
        
        & & & & &Default & Calibrated & Default & Calibrated \\
        \midrule
        adder\_n4 & 4 &16 &10           & 41 & 0.13 
        &0.10  &0.87 & 0.90\\
        adder\_n10 & 10 & 146 & 136     & 503 & 0.56 
        &0.49   &0.44   &0.51\\
        dnn\_n8   & 8 & 237 & 63        & 249 & 0.42
        &0.37  & 0.71 & 0.76\\
        ising\_n10 & 10 & 90 & 20       & 111 & 0.32
        &0.24  & 0.76 &0.84 \\
        qpe\_n9 & 9 & 97 & 86           & 339 & 0.15
        & 0.07 &0.94 & 0.98\\
        cat\_state\_n22 & 22 & 21 & 21  & 66 & 0.39
        & 0.36  & 0.61 & 0.64\\
        ghz\_state\_n23 & 23 & 22 & 22 & 65 & 0.47
        & 0.41 & 0.53 & 0.59\\
        qram\_n20 & 20 & 352 & 225 & 794 & 0.74
        & 0.68  & 0.26 & 0.32\\
    \bottomrule
    \end{tabular}
    \caption{Application-level benchmarking results for ibm\_rensselaer.}
    \label{tab:application}
\end{table*}
\subsection{Calibration-level Benchmarking}
\label{improvementincalibrationruntime}
Leveraging parallel graph traversal in the calibration process brings about a huge opportunity for acceleration. During experiments on real quantum machines, it was observed that when the current calibration subgraph—representing all calibration tasks executed simultaneously—includes a large number of complex custom pulses defined by arrays, the parallel execution often results in errors. Therefore, calibration subgraphs that contain more than 20 edges are split into smaller graphs with no more than 10 edges each. When splitting the edges, edges that are scheduled for Direct CR calibration are separated from edges scheduled with other calibration methods. Therefore, the long calibration time of Direct CR does not interfere with the runtime of other subgraphs. The calibration subgraphs are then executed sequentially as previously mentioned. As shown in Figure~\ref{fig:calibrationtime}, even limited by the hardware constraints, the parallel calibration could reduce the total runtime by a factor of 7.9. Given an ideal hardware, the improvement in runtime could be as large as 25 times.

\subsection{Device-level Benchmarking}
\label{devicebenchmarkingresults}
\textbf{Quantum Volume.} Quantum Volume is a single-number metric utilized to characterize the power of a quantum computer. The circuit used to measure Quantum Volume is composed of random unitaries involving $d$ qubits and has a depth of $d$. Beginning with a small $d$, the Quantum Volume is measured with repetitive trails. Once the measurement outcome satisfies the fidelity requirement, the experiment proceeds with a larger $d$. The fidelity requirement is shown below
\begin{equation}
    \frac{n_h - 2 \sqrt{n_h(n_s - \frac{n_h}{n_c})}}{n_c n_s} > \frac{2}{3}
\end{equation}
where $n_h$ is the number of heavy outputs (outputs with probabilities higher than the median probability), $n_c$ is the number of circuits created which is set to be 100, and  $n_s$ is the number of shots. Such a threshold checks if there is at least a 97\% chance the heavy output probability is greater than 2/3. Finally, until a square circuit with $d$ depth cannot pass the fidelity requirement, the Quantum Volume is represented as $2^d$.\\
\textbf{Error Per Layered Gate (EPLG).} As current quantum hardware scales to over 127 qubits, Quantum Volume is still a convincing holistic test of processor performance. However, it only contains $d$ qubits with the highest fidelity. The layer fidelity fills the void as the protocol involves as many connected qubits as required. The layer fidelity is measured by splitting selected $n$ qubits into $M$ disjoint layers where each layer contains non-overlapping two-qubit gates and idle qubits. Simultaneous direct randomized benchmarking is conducted on the disjoint layer to measure the error of two-qubit gates and idle qubits. A process fidelity $F_i = \frac{1 + (d^2 -1)\alpha}{d^2}$ is obtained for each measured decay where $d = 2^{n}$, and $\alpha$ is the decay rate. The layer fidelity and EPLG is calculated as
\begin{align}
    LF = \prod_{m}^{M}\prod_{j}F_{j,m} \\
    EPLG = 1 - LF^{1/n}
\end{align}

The device-level benchmarking results for ibm\_rensselaer and ibm\_nazca are shown in Table~\ref{tab:devicebenchmarking}. For both machines, the Quantum Volume is enlarged from 128 to 256, indicating that the quantum processors are able to execute a square circuit with a width of 8 and a depth of 8. Meanwhile, the EPLG for ibm\_rensselaer is reduced by a factor of 2.3 while EPLG for ibm\_nazca also achieves a reduction of 1.99 times, proving the overall improvement in two-qubit gate fidelity. From both local metric (QV) and global metric (EPLG), our optimal pulse profiling technique exhibits great potential in further elevating the performance of current quantum hardware.

\subsection{Application-level Benchmarking}
To evaluate the real-world effect brought about by the advanced optimal pulse profiling technique, several quantum programs from OpenQASMBench are adopted to examine the application-level performance. Through running programs on noise-free simulators, quantum computer with default pulse setup, and quantum computer with optimal pulse for each qubit pair, the error rate (E) and the fidelity (F) of each circuit are obtained as
\begin{align}
E &= \frac{1}{2} \sum_{x} \left| P_{\text{ideal}}(x) - P_{\text{real}}(x) \right| \\
F &= \left( \sum_{x} \sqrt{P_{\text{ideal}}(x) \cdot P_{\text{real}}(x)} \right)^2
\end{align}
where $P_{\text{ideal}}(x)$ represents the ideal probability of outcome $x$ and $P_{\text{real}}(x)$ represents the actual probability of outcome $x$ from real quantum hardware.  As demonstrated in Table~\ref{tab:application}, our optimal pulse profiling technique exhibits error rate reduction and fidelity increase in all benchmarks, with a maximum fidelity increase of 16\%, proving its enormous potential in leveraging the hidden power of current quantum hardware.

\section{Related Works}
\label{relatedworks}
Recent work on optimizing quantum circuits for higher Quantum Volume underscores the importance of gate speed in circuit performance~\cite{jurcevic2021demonstration}. Researchers have developed techniques to reduce two-qubit gate durations by incorporating pre- and post-single-qubit rotations into the compilation, effectively shortening the gate to its entangling component. Additionally, high-fidelity alternatives to traditional echo pulse sequences, like direct echo-free CX gates leveraging target rotary pulsing~\cite{sundaresan2020reducing}, have been proposed to eliminate extra single-qubit gates from the duration. These approaches demonstrate that a strategic balance between gate speed and fidelity can enhance overall circuit performance, however, this approach much more expensive than the ECR gate calibration in terms of the calibration cost.

And previous studies on CR gates have demonstrated significant advancements in mitigating control errors and achieving high-fidelity quantum operations. ~\cite{Li2024Experimental} inspired by the DRAG framework, introduces recursive corrections to mitigate single- and multi-photon transition errors, achieving improvement over standard calibration techniques, ~\cite{chow2011simple} leverages amplitude control of a microwave drive on one qubit at the resonance frequency of another to generate the desired interaction. However, those method overlooked the limitation on hardware inconsistencies and directly applies the same pulse envelope to all pairs of qubits.

\section{Conclusion}
\label{conclusion}
In this paper, we present a fine-grained calibration protocol that addresses both hardware variations and parallel execution needs in quantum computers. Our protocol combines three hardware-aware calibration policies with a graph-based method for parallel calibration. Experiments on quantum processors with up to 127 qubits demonstrate significant fidelity improvements while maintaining practical calibration overhead. This work advances hardware-aware calibration strategies and establishes a foundation for fault-tolerant quantum computing at scale.

\section{Acknowledgement}
We acknowledge the usage of IBM Quantum Services for this work.

\newpage
\bibliographystyle{plain}
\bibliography{refs}
\end{document}